\documentclass[preprint,12pt]{aastex}
\usepackage{emulateapj5}
\newcommand{\degree}{\mbox{$^{\circ}$}}
\newcommand{\am}{\mbox{\arcmin}}
\newcommand{\as}{\mbox{\arcsec}}

\def\lsim {$\rlap{\raise.4ex\hbox{$<$}}\lower.55ex\hbox{$\sim$}\,$}

\newcommand{\dcop}{DCO$^+$}

\input{epsf}

\def\c17o{$\rm C^{17}O$}
\def\dc18o{$\rm C^{18}O$}
\def\dcop{$\rm DCO^{+}$}
\def\h13cop{$\rm H^{13}CO^{+}$}

\begin{document}

\title {\bf Chemistry and Dynamics in Pre-Protostellar Cores} 
\author {Jeong-Eun Lee, Neal J. Evans II, and Yancy L. Shirley}
\affil{\it Department of Astronomy, The University of Texas at Austin,
       Austin, Texas 78712--1083}
\email{jelee, nje@astro.as.utexas.edu, yshirley@aoc.nrao.edu}
\and
\author {Ken'ichi Tatematsu}
\affil{\it National Astronomical Observatory of Japan, 2-21-1 Osawa, Mitaka,
Tokyo, 181-8588, Japan}
\email{k.tatematsu@nao.ac.jp}

\begin{abstract}

We have compared molecular line emission to dust continuum emission and 
modeled molecular lines using Monte Carlo simulations in order to study the
depletion of molecules and the ionization fraction in three 
preprotostellar cores, L1512, L1544, and L1689B.
L1512 is much less dense than L1544 and L1689B, which have similar density 
structures. L1689B has a different environment from those of L1512 and L1544.
We used density and temperature profiles, calculated
by modeling dust continuum emission in the submillimeter, 
for modeling molecular line profiles.
In addition, we have used molecular line profiles and maps observed in 
several different molecules toward the three cores.
We find a considerable diversity in chemical state among the three cores.
The molecules include those sensitive to different timescales of chemical 
evolution such as CCS, the isotopes of CO and $\rm HCO^+$, $\rm DCO^+$, 
and $\rm N_2H^+$. 
The CO molecule is significantly depleted in L1512 
and L1544, but not in L1689B.
CCS may be in the second enhancement of its abundance in L1512 and L1544 because
of the significant depletion of CO molecules. $\rm N_2H^+$ might already start 
to be depleted in L1512, but it traces very well the distribution of dust 
emission in L1544. On the other hand, L1689B may be so young that 
$\rm N_2H^+$ has not reached its maximum yet.
The ionization fraction has been calculated using 
$\rm H^{13}CO^+$ and $\rm DCO^+$. The result shows that the ionization fraction
is similar toward the centers of the three cores. 
This study suggests that chemical evolution depends on the absolute 
timescale during which a core stays in a given environment as well as its 
density structure. 

\end{abstract}

\section{INTRODUCTION}

Many studies have been done on low mass-star formation in detail, 
theoretically and observationally, over the past two decades. 
These studies were possible because many low mass protostellar cores are close, 
isolated, and geometrically simple. In addition, they evolve slowly, allowing 
the process to be observed in distinct stages.  
The information accumulated by those studies now provides the big picture 
of the evolutionary stages of low-mass star formation. 
Lada (1987), Myers \& Ladd (1993) and Andr\'e et al. (1993) developed an 
evolutionary sequence based on the spectral energy distribution (SED)
of dust continuum emission through Class 0 to Class III sources, 
and Adams et al. (1987) supported the evolutionary sequence by modeling 
the SED theoretically. 
In addition, several groups (Larson 1969, Penston 1969, Shu 1977, Shu et al. 
1987, Foster \& Chevalier 1993, Ciolek \& Mouschovias 1994, McLaughlin \& 
Pudritz 1997) have developed dynamical evolution models that theoretically 
predict different structures of density and velocity in protostellar cores,
based on differing initial conditions.

In spite of the big picture of low-mass star formation, 
the initial collapse is poorly understood. 
The collapse dynamics and the timescale of evolution are crucially 
dependent on initial conditions. 
Preprotostellar cores (hereafter PPCs) are believed to be gravitationally
bound, but they have no central hydrostatic protostar 
(Ward-Thompson et al. 1994, 1999). 
Therefore, PPCs can be considered as the potential sites of future star
formation and give us chances to probe the initial conditions of star 
formation. 
The series of papers of Ward-Thompson et al. (1994, 1999, 2000, and 2001) 
and the paper of Andr\'e et al. (1996) argued that the density of 
PPCs is relatively constant in the inner cores and falls off at larger radii.
This is very different from the singular isothermal sphere that leads to the 
``inside-out" collapse (Shu 1977). 
However, they assumed a uniform temperature in PPCs to calculate density 
structures. Evans et al. (2001) and Zucconi et al. (2001) show that the 
temperature of a dust grain decreases toward the centers of PPCs, assuming that
dust grains are heated only by the attenuated external radiation field 
and that the radiation from the core itself is optically thin. 
Based on their results, Evans et al. (2001) argue that the continuum 
emission is significantly dependent on the temperature structure as well as
the density structure, and they show that Bonnor-Ebert spheres can fit well
the submillimeter dust continuum emission in PPCs.
In agreement, the recent paper of Ward-Thompson et al. (2002) has shown that
the color temperatures of PPCs that are calculated from the 170 and 200 
$\rm \mu m$ dust emission observed by ISOPHOT decrease toward their centers. 
As a result of the lower temperature at the center, the density at the center
must be more peaked than is the intensity of dust continuum emission.
Evans et al. (2001) showed that even a power law could fit the density structure
of L1544.

Ciolek \& Mouschovias (1994) modeled ambipolar diffusion for the 
quasi-static evolution of a uniformly magnetized isothermal molecular cloud,
finding relatively constant density toward the center.
However, there has been debate about the importance of ambipolar diffusion
because of differences between inward velocity observed 
(Tafalla et al. 1998, Williams et al. 1999) and calculated from the 
magnetic field (Goodman and Heiles 1994, Andr\'e et al. 1996, 
Ward-Thompson et al. 1999, Crutcher and Troland 2000,
Ward-Thompson et al. 2000). 
Those differences might be resolved  
by calculating more accurate density structure 
and, in turn, the correct temperature structure from the observed data.  
Ciolek and Basu (2000) showed if the density at the center of a PPC were 
higher, then their model could explain the observed magnetic field and 
inward velocity.

In addition to density and temperature distributions, we need to understand
the ionization structure because magnetic relaxation by ambipolar diffusion 
is tightly related to the ionization structure (Rawlings 2000).
Ionization fraction can affect the process of gravitational
collapse because the coupling between ions and magnetic field
plays a key role in preventing free-fall collapse of a dense core
and slowing the drift of neutral gas through the magnetic field 
(ambipolar diffusion).
In addition, the gas-phase chemical reactions depend significantly on ionization
fraction, and the ionization fraction is highly sensitive to the depletion of 
molecules because some molecules play a crucial role in destroying ions.
The degree of ionization in cold dense cores can be measured by the 
$\rm [DCO^+]/[HCO^+]$ abundance ratio (Watson 1977), which is 
directly related to the electron density and depletion fraction.
Recently, many molecular line observations show evidence for chemical 
differentiation and the depletion of some molecules such as $\rm C^{17}O$, 
$\rm C^{18}O$, and CCS in dense cores 
(Lada et al. 1994, Kuiper et al. 1996, Black et al. 1995, McMullin et al. 1994, 
Benson and Myers 1998, Willacy et al. 1998, Kramer et al. 1999, 
Ohashi et al. 1999, Caselli et al. 1999, Jessop and Ward-Thompson 2001, 
Tafalla et al. 2002).  
They suggested that this trend is due to
the substantial depletion of molecules onto dust grain surfaces.
In addition, several groups (Bergin and Langer 1997, Aikawa et al. 2001,
Caselli et al. 2002, Li et al. 2002) modeled chemical evolution, 
including gas-dust interaction as well as gas-gas interaction, 
to show the changing depletions of some molecules and ionization fraction with time 
or density, and most of them compared their results to real observations. 
Therefore, both gas phase chemical processes and dust-gas interactions 
should be considered in studying star formation.

Those theoretical and observational results suggest that the relative 
distributions of several molecules can be used to probe the timescale of the 
dynamical evolution of a PPC.
However, Rawlings et al. (1992) and Rawlings and Yates (2001) modeled 
several molecular lines combining self-consistently dynamical, chemical, 
and radiative transfer models, and showed that the abundance variation 
and the line profiles are very sensitive to the free parameters in the 
chemical model. Consequently, we need to use multiple transitions and  
maps in order to constrain these models.  

In this paper, we explore the chemical and dynamical differences in three 
PPCs (L1512, L1544, and L1689B) comparing molecular depletion and
ionization fraction by simple analysis of integrated intensity and column 
density and by modeling several molecular line profiles in detail. 
In this study, we cover a very early-time molecule (CCS), an early-time 
molecule (CO), a middle-time molecule ($\rm HCO^+$), and a very 
late-time molecule ($\rm N_2H^+$) in chemical evolution. 
According to the chemical models of Bergin and Langer (1997) and 
Aikawa et al. (2001) that used an initial density of $10^4$ cm$^{-3}$,
the CCS abundance decreases 
rapidly after a few times $10^5$ years since CCS is a polar molecule, and 
it is easily frozen out onto the surface of dust grains. 
In addition, CCS is destroyed by HCO$^+$ or H$_3$O$^+$, whose 
abundances increase until about $10^6$ years.
Even though CO is a non-polar molecule, it has very high binding energy (1740 K
onto water ice mantle grains, Bergin and Langer 1997). 
As a result, CO starts to be depleted onto the surface of dust grains after
about $10^6$ years causing the depletion of HCO$^+$ in turn.
HCO$^+$ is formed by the interaction between H$_3^+$ and CO or H$_2$D$^+$ and
CO. However, N$_2$, which is the precursor of N$_2$H$^+$, starts to be abundant 
much later than other molecules and reaches its maximum around $10^7$ years. 
N$_2$ has low binding energy (750 K, Aikawa et al. 2001) onto the surface of dust
grains, so it is not frozen out from the gas phase easily. 
In addition, H$_3^+$, which is the 
other precursor of N$_2$H$^+$, increases as CO is depleted. Another cause of
the enhancement of N$_2$H$^+$ at late times is the decline of the destruction
of N$_2$H$^+$ by CO, which is significantly depleted at late times.

We observed multiple positions in each core to study the chemical 
distribution within a core  
in addition to studying the chemical differences among the three PPCs.
The most important aspect of this study is that we start with reasonable 
physical models calculated from dust analysis. 
The density and temperature structures found by Evans et al. (2001) 
are adopted for modeling molecular line profiles.  

In section 2, we summarize observational parameters, and section 3 shows
the results of observations. In section 4, we analyze column densities of 
several molecules. Section 5 compares modeled
molecular line profiles and observed line profiles. Finally, we discuss 
our results regarding depletion and ionization in the view of chemical and 
dynamical evolution in section 6.

\section{OBSERVATION}

We observed the three pre-protostellar cores 
in \dc18o ($\rm J=2-1$ and $\rm J=3-2$), \c17o ($\rm J=2-1$), \dcop 
($\rm J=3-2$), $\rm HCO^+$ ($\rm J=3-2$), and
$\rm H^{13}CO^{+}$ ($\rm J=3-2$) with the 10.4 m
telescope of the Caltech Submillimeter Observatory (CSO) at Mauna Kea, 
Hawaii from 1995 to 2002. 
We used an SIS receiver with an 
acousto-optic spectrometer (AOS) with a 50 MHz bandwidth and 1024 channels, 
and the frequency resolution was about 2 to 2.5 channels, 0.13 to 0.16 $\rm km~s^{-1}$ 
at around 220 GHz. The pointing uncertainty was about 4$\arcsec$ on 
average.  
We also observed these cores in $\rm H^{13}CO^{+}$ ($\rm J=1-0$), $\rm N_2H^+$ 
($\rm J=1-0$), and CCS ($\rm N_J=4_3-3_2$) with the 45 m telescope
of the Nobeyama Radio Observatory in Japan on January 2002.
We used the S40, S80, and S100 receivers with an AOS with
40 MHz bandwidth and 2048 channels.
The frequency resolution was about 2 channels (about 40 kHz) so that the 
velocity resolution was about 0.27, 0.14, and 0.13 km s$^{-1}$ at 43, 86, 
and 100 GHz, respectively.
The observed sources are listed in Table 1; the observed lines and the main
beam efficiency ($\eta_{mb}$) of each observation are summarized in Table 2.
In Table 1, the last column shows the offsets of the dust peaks from the given
coordinates; all offsets in figures other than Figure 2 and 3 are relative to
the dust peak, assumed to be the center of the core in all modeling.

\section{RESULTS}

\subsection{Spectra}

Figure 1 shows the spectra of molecular lines observed toward the centers of
the three PPCs. The centers are the positions of the centroid of the dust 
emission (Shirley et al. 2000). 
Most of the molecular lines have not been mapped completely
but observed through two orthogonal cuts in $\alpha-\delta$, crossing at 
the centers of dust emission in the cores.
We do not see any consistent trend in the intensity of each spectrum 
from core to core. For example, \dc18o and \c17o lines are the strongest 
in L1689B, but $\rm N_2H^+$ and CCS are the strongest in L1544. 
All the lines except for $\rm N_2H^+$ and CCS are much weaker in L1512
than in L1689B. 
The line width (FWHM) of the \dc18o $\rm J=2-1$ line is about 0.3, 0.5, and 
0.6 km s$^{-1}$ in L1512, L1544, and L1689B, respectively.

\subsection{Maps}

Figure 2 shows maps of three PPCs, where the grey scale indicates 850 
$\rm \mu m$ dust continuum intensity and the contours are for integrated 
intensities of $\rm C^{18}O$ $\rm J=2-1$ and $\rm DCO^+$ $\rm J=3-2$ lines.
We can see that the peaks of dust emission and $\rm C^{18}O$ $\rm J=2-1$ 
emission are not consistent. In the map for L1512, especially, the centroid of
the dust emission coincides with a hole in the $\rm C^{18}O$ $\rm J=2-1$ 
emission. In contrast, the
distribution of $\rm DCO^+$ $\rm J=3-2$ emission appears to peak toward the 
center of dust emission in L1544 and L1689B, even though our maps of 
$\rm DCO^+$ $\rm J=3-2$ emission are not complete.

\section{SIMPLE ANALYSIS}

In this section, we use a simple analysis with standard techniques for 
comparison to similar work done by others, and, in the next section, we will
deal with detailed models.
We use dust emission as the tracer of column density least affected 
by chemistry in this section, and we assume that the dust emissivities do not 
change with radius.
If temperature decreases toward the center, the column density 
calculated from dust emission with the assumption of an 
isothermal case could be underestimated toward the center.
Even though the absolute column density calculated from the dust emission can 
change with the assumed dust opacity by a factor of 2 or 3, its relative 
distribution with radius can be traced relatively well by dust emission.   

\subsection{Integrated Intensities}

We can compare the distribution of a molecular component with that of the 
dust component
within a core using the integrated intensity of an optically thin molecular 
line and the dust continuum intensity observed at a submillimeter wavelength.  
In the upper panels of Figure 3, we plot the logarithmic integrated intensities 
of CCS $\rm N_J=4_3-3_2$, H$^{13}$CO$^+$ $\rm J=1-0$, and $\rm N_2H^+$ $\rm 
J=1-0$ lines in order to compare to the 850 $\rm \mu m$ dust continuum intensity
through the cuts with constant declination shown in Figure 2.
The dust continuum intensity and the H$^{13}$CO$^+$ intensity have been shifted
upward to compare the shapes of intensity distributions with radius more 
effectively, and the intensity of dust continuum has been measured into 
apertures of 20$\arcsec$, spaced by 20$\arcsec$ for consistency with 
the H$^{13}$CO$^+$ and $\rm N_2H^+$ observations. 

The dust emission of L1512 is about half of that of L1544, and the distribution
is not as peaked as that of L1544 and L1689B.
L1544 and L1689B have similar dust emission even though it is more peaked in 
L1544. However, the CCS intensity of L1689B is more similar to that of L1512 
rather than to
that of L1544, and the CCS intensity of L1544 is about twice those in L1512 and
L1689B. The CCS intensity in all the three PPCs is not peaked as much toward 
the center as the dust emission. 
However, it has a bump toward the centers of
L1512 and L1544 (see \S 6.1). In L1689B, the bump is not certain because
of the low signal-to-noise ratio.

If we check the relative variations of the integrated intensity of
the $\rm H^{13}CO^+$ $\rm J=1-0$ line with radius in each core,
in L1689B, it is skewed to the west but follows well the dust
continuum emission, decreasing at the edge of the core.
On the other hand, the distribution of the integrated intensity of 
the $\rm H^{13}CO^+$ $\rm J=1-0$ line in L1512 is not well matched with 
dust continuum, and the integrated intensity is even depressed toward
the center.
We also see the depression of $\rm H^{13}CO^+$ emission at the center of L1544.
This depression was predicted by Caselli et al. (2002) in their chemical 
models, but not observed, probably because their angular resolution 
(28$\arcsec$) was worse than ours (18$\arcsec$).  
We suspect that the optical depth of the $\rm H^{13}CO^+$ 
$\rm J=1-0$ line in L1544 and L1689B is not negligible.
Therefore, the depression in those cores could be partially caused by 
optical depth. 
However, L1689B does not show a depression even though it has a density 
structure similar to L1544 suggesting that depletion is playing a major role
in L1544.

The $\rm N_2H^+$ intensity is well peaked in L1544 and shows almost the same 
shape as the dust emission. L1689B also shows similar shapes 
in dust emission and $\rm N_2H^+$ intensity.
However, the $\rm N_2H^+$ intensity of L1689B is only two thirds of that of 
L1544, even though these two cores have very similar dust emission and density
structures, with central densities of $10^6$ cm$^{-3}$.
In L1689B, the peak of the intensity of $\rm N_2H^+$ is shifted from that of 
the intensity of the dust emission similarly to the intensity of 
$\rm H^{13}CO^+$. 
However, the ratio of the intensity of $\rm H^{13}CO^+$ to the intensity of 
$\rm N_2H^+$ at the center of L1689B, is greater than that in the other two 
cores. 

In L1512, with a central density of $10^5$ cm$^{-3}$, 
$\rm N_2H^+$ emission is more extended than the dust, unlike in the other cores.
Since the optical depth of the $\rm N_2H^+$ $\rm J=1-0$ line in L1512 
is smaller than in L1544 (Benson et al. 1998), 
this might indicate the possible presence of depletion at the dust peak.

In the following section, 
we compare the distributions of CO isotopes to that of 
dust emission by using molecular hydrogen column densities calculated from them
instead of using the intensity distribution because the 
optical depth effect in the C$^{17}$O line can be calculated, and the standard
abundance of CO is known.  

\subsection{Column Density}

We compare the $\rm H_2$ column densities 
calculated from observed molecular line intensities with that 
calculated from submillimeter continuum emission.
This simple analysis assumes that 1) dust continuum emission traces all 
material along a line of sight in a core, and dust and gas are well mixed,
2) the core is isothermal, and dust and gas are well coupled so that 
they have same temperature, 3) molecular abundances are constant through 
the core, and 4) all levels are in LTE. Therefore, the excitation
temperatures in all levels are the same as the kinetic temperature and
the same as the dust temperature.

\subsubsection {The ${\rm H_2}$ column density from molecular emission lines}

If we assume a line is very optically thin so that absorption can be completely
neglected, the equation of radiative transfer is given by

\begin{equation}
\frac{dI_\nu}{ds}=\frac{h\nu_{ul}A_{ul}}{4\pi}n_u,
\end{equation}

\noindent where $n_u$ is the number density of upper level, and $A_{ul}$ is the 
Einstein coefficient.
For the transition $\rm J\rightarrow J-1$ in a linear molecule, the resulting
relation between $N(x)$, the column density of molecule $x$, and the integrated
intensity of the line, is

\begin{equation}
N(x)=\frac{3kQe^{E_u/kT_{ex}}}{8\pi^3\nu\mu^2J}{\int T_R~dv},
\end{equation}

\noindent where $E_u=hBJ(J+1)$, $B$ is the rotational constant,
$T_{ex}$ is the excitation temperature above the ground state,
$\mu$ is the dipole moment, and $Q$ is the partition function. 
Equation 2 is valid only in the limit that $\tau \rightarrow 0$.
For finite optical depth, an optical depth correction can be applied:

\begin{equation}
N_{thick}=N_{thin} \frac{\tau_\nu}{1-e^{-\tau_\nu}}.
\end{equation}

In principle, one should also correct for the presence of the cosmic background
radiation, $T_{cmb}$, but for the lines considered here, at $\lambda < 1.4$ mm,
the Planck function in temperature units is less than 0.23 K, completely
negligible compared to likely values of $T_{ex}$. The effect on $N(x)$ of
neglecting $T_{cmb}$ is less than 5\% for $T_{ex} \geq 10$ K and $\tau \leq 1$,
typical of the conditions relevant here. These effects 
are less than any plausible calibration uncertainty.

The rotational constant and the dipole moment of
each molecule used for this calculation are listed in Table 3. 
For the excitation temperature, we used the kinetic 
temperature of 10 K, a typical value in dark molecular 
clouds, under the assumption that all lines are thermalized, that is, 
$T_i=T_{ex}=T_k=10$ K.
If we use 20 K for this calculation, the result is changed only by a 
factor of 1.2 for the $\rm J=2-1$ line of \dc18o and \c17o.

We calculated the column densities using $\rm C^{18}O$ $\rm J=2-1$ and
$\rm C^{17}O$ $\rm J=2-1$ lines.
First, the lines were assumed to be optically thin.
However, we can calculate the optical depth of the $\rm C^{17}O$ $\rm J=2-1$ 
line directly by comparing the relative intensities of its 9 hyperfine 
components (Ladd et al. 1998).
The optical depth of the $\rm C^{17}O$ $\rm J=2-1$ line was obtained by a 
fitting method in the program CLASS. 
The panels in the third row of Figure 1 shows the results of fitting the
hyperfine components of the $\rm C^{17}O$ $\rm J=2-1$ line in the center of 
each core.  
The total optical depths for L1544 and L1689B are not small enough to be ignored
(Table 4), even though each individual hyperfine component is optically thin.
In order to correct for this optical depth effect, we applied equation (3) to 
the above calculation for an optically thin line. 
For the \c17o\ $J = 2-1$ line, 
$\tau_\nu$ is the optical depth of the line center that is calculated by
multiplying the relative strength (0.693) of the main group, which has four 
close components, by the total optical depth. 
The correction factors at the centers of L1544 and L1689B are about 1.4. 
Note that we do not use the excitation temperature of \c17o $\rm J=2-1$ 
obtained by fitting the hyperfine structure to calculate the column
density of \c17o because this method assumes that $T_{ex}$ is constant along
the line of sight. Our detailed models (see \S 5.3.1) shows that this 
assumption leads to incorrect results. Also we want to use the same method 
for \dc18o and \c17o and compare the results.

Finally, we convert the column density of each molecule into the $\rm H_2$
column density using the abundance ($X$) of molecule $x$;

\begin{equation}
N({\rm H_2})=\frac{N(x)}{X(x)}~~~~{\rm cm^{-2}}.
\end{equation}

\noindent We used a CO abundance of $2.7\times10^{-4}$ (Lacy et al. 1994), 
about three times bigger than the abundance used in other studies such as 
Bacmann et al. (2002) and Jorgensen et al. (2002).
The abundance ratios [CO/\dc18o] and [\dc18o/\c17o] are about 560 ($\pm$ 25)
and 3.2 ($\pm$ 0.2), respectively (Wilson and Rood, 1994).
The abundance of each molecule used in calculating the $\rm H_2$ column density
is listed in Table 3.

\subsubsection{The ${\rm H_2}$ column density from dust continuum emission}

If the emission from dust is optically thin, the observed flux density 
($S_{\nu}$) can be related to the column density of gas by

\begin{equation}
N({\rm H_2})=\frac{S_{\nu}}{\mu m_H \kappa_\nu B_\nu(T_d) \Omega},
\end{equation}
where 
$\mu$ is the mean molecular weight,
$m_H$ is the atomic mass unit,
$\kappa_\nu$ is the mass opacity of dust per gram of gas, 
$B_{\nu}$ is the Planck function, 
and $\Omega$ is the aperture solid angle.
$\Omega=(\pi \theta^2)/(4 ln 2)$ for a circular Gaussian aperture.

We use column 5 of the Ossenkopf-Henning dust opacity table
(Ossenkopf \& Henning 1994), which represents
agglomerated dust grains with thin ice mantles, and assume a gas-to-dust mass 
ratio of 100 to get $\kappa_{850}=0.018 ~~\rm cm^2~g^{-1}$. 
We use the 850 \micron\ data from Shirley et al. (2000) but measure
the flux in an aperture of 33$\arcsec$, 
the size of main beam in the observations of C$^{17}$O. 
If the dust temperature, $T_d$ is 10 K, the $\rm H_2$ 
column density can be calculated from

\begin{equation}
N(H_2)=3.36\times 10^{22} \times S_{850}({\rm Jy})~~~~{\rm cm^{-2}}.
\end{equation}

\noindent If the temperature were 20 K, the calculated $\rm N(H_2)$
would be lower by a factor of 3.3. 
 
\subsubsection{Results}

$\rm C^{18}O$ $\rm J=2-1$ and $\rm C^{17}O$ $\rm J=2-1$ lines have been used 
to calculate the ${\rm H_2}$ column density assuming constant molecular 
abundances.
Basically, we assume the lines are optically thin. 
If a molecular line is optically thick, then the calculation with
equation (2) and (4) underestimates the ${\rm H_2}$ column density.
As mentioned in the previous section, however, we can calculate the optical 
depth of the $\rm C^{17}O$ $\rm J=2-1$ line by fitting its hyperfine structure.
We have corrected the column density of the  $\rm C^{17}O$ $\rm J=2-1$ line
for optical depth in L1544 and L1689B even though the correction is not very 
big. 

The results of fitting the hyperfine components of \c17o $\rm J=2-1$ show that
the total optical depth of this line is about 0.1 in L1512, but it is not
negligible at the central positions in L1544 ($\tau \approx 1.0$) and L1689B 
($\tau \approx 0.8$). In L1512, the error of the calculated optical depth is 
very big, because of the low S/N in the weak components.
However, the ratio ($\approx$ 4.3) between the strongest component and the 
second strongest component in L1512 is bigger than that ($\approx$ 3.3) in the 
two other cores, so the total
optical depth would be less than that of L1689B, at least.
The abundance of \dc18o is about 3.2 times greater than that of \c17o.
Therefore, the optical depth of \dc18o is about 3.2 greater than the total
optical depth of \c17o, and the optical depth of \dc18o $\rm J=2-1$ must be 
considered in calculating column densities.
We plot the ${\rm H_2}$ column density calculated from the \c17o $\rm J=2-1$ 
line in the lower panes of Figure 3, where the optical depth effect has 
been corrected by equation (3).
However, in the figure, we use the ${\rm H_2}$ column density calculated from 
the \dc18o $\rm J=2-1$ line based on the optically thin line approximation 
without correcting for its optical depth in order to show
the significance of the optical depth effect.  
In fact, \c17o has been observed in fewer positions so that we cannot calculate
the optical depth of every \dc18o spectrum.

In the figure, we see significant differences between the ${\rm H_2}$ column 
densities calculated from the 850 $\rm \mu m$ emission and 
the \dc18o $\rm J=2-1$ line in all three PPCs.
At central positions, the ${\rm H_2}$ column density calculated from the \dc18o 
$\rm J=2-1$ line is about 15 times smaller in L1512 and L1544,
and 8 times smaller in L1689B than the ${\rm H_2}$ column
densities calculated from the $\rm S_{850}$ $\rm \mu m$ emission.
However, after correcting for the optical depth of the \c17o $\rm J=2-1$ line, 
the difference between the ${\rm H_2}$ column densities calculated from 
$\rm S_{850}$ and from \c17o has been reduced. 
In L1689B, especially, the difference is just a factor of 2 after the 
correction.

Another interesting thing is that the distribution of \c17o in L1544 does not 
show the central depression observed in Caselli et al. (1999). This is 
probably due to the larger beam (33\arcsec) in this observation compared to
that (21\arcsec) of Caselli et al. (1999), similarly to the case of 
H$^{13}$CO$^+$ (see \S 4.1). This result suggests that inadequate 
resolution can prevent us for calculating the actual depletion of molecules.  

\subsection{Conclusions}

The integrated intensities and the ${\rm H_2}$ column densities calculated from 
several molecules have been compared with the intensity of 
850 $\rm \mu m$ dust emission and the ${\rm H_2}$ column densities
calculated from the dust emission in our simple analysis of molecular line 
observations. The first result of this analysis is that the dust emission is
more peaked than the molecular emission. The depression of molecular emission 
toward the centers is partially caused by the optical depth effect on molecular
lines. We could see this effect comparing the column densities calculated from 
\c17o $\rm J=2-1$ (marginally optically thin and corrected by $\tau$) and 
\dc18o $\rm J=2-1$ (optically thick but not corrected by $\tau$) lines. 
The correction for the optical depth effect by the \c17o $\rm J=2-1$ line
removes some of the difference between the ${\rm H_2}$ column densities
calculated from the \dc18o $\rm J=2-1$ line and dust emission. The remaining 
considerable discrepancy would be caused by the depletion of CO molecules.  
This result shows that we can overestimate the depletion of a molecule if we 
use an optically thick line without correcting for the optical depth.
Jorgensen et al. (2002) have compared the H$_2$ column densities calculated 
from the 1.3 mm dust continuum and the C$^{17}$O 1$-$0 line and found similar 
depletion factors of CO to ours in L1544 and L1689B.
We will explore the depletion in detail through modeling molecular line 
profiles using the Monte Carlo method in the next section. 
We emphasize that the amount of the discrepancy varies from core
to core. After the correction for the optical depth effect, the  ${\rm H_2}$
column density calculated from the \c17o $\rm J=2-1$ line is 11 in L1512, 9 in 
L1544, and 2 in L1689B times smaller than the  ${\rm H_2}$ column density
calculated from dust emission (Table 5). 

The simple method used in this analysis clearly reveals evidence for depletion,
but it has several limitations for quantitative analysis. First, the
beam size of each line observation is not the same. 
If we had mapped these cores 
completely in each line, then we would have been able to convolve the maps 
using the same beam size. 
Second, the variations of the excitation temperature and the abundance of a 
molecule along the line of sight are not included in this simple analysis. 
In addition, the 
absolute ${\rm H_2}$ column density calculated from dust emission may be 
uncertain by a factor of 2 to 3 because of the uncertainty of dust opacity. 
Due to these limitations, we have to model molecular lines and compare them 
to the observed lines in order to study depletion more in detail. 

\section{DETAILED MODELS}

\subsection{Method}

We used the Monte Carlo (hereafter MC) method to calculate the radiative 
transfer 
of molecular lines (Bernes 1979, Choi et al. 1995). The MC code for this work
has been developed by Choi et al. (1995). The MC code generates model photons 
at a random position, in a random direction, and at a random frequency with 
proper random number distributions. These photons go through a one-dimensional 
spherically symmetric molecular cloud adjusting the level populations of 
molecules. The MC code calculates the excitation by the model photons and uses
statistical equilibrium to adjust each level population until the criteria
of convergence are satisfied.
The MC method is more powerful than the LVG method because it 
can deal with arbitrary distributions of systematic velocity, density, 
kinetic temperature, microturbulence, and abundance self-consistently.
Therefore, we can combine reasonable physical models from analysis of dust 
emission with chemical models to produce molecular line simulations, 
and the depletion of molecules with radius can be tested. 

For the MC simulation, we needed collision rates at lower gas temperatures than
are usually available, so we extrapolated linearly to 5 K the collision 
rates for CO (Flower and Launay 1985), which we used for \dc18o and \c17o. 
Similarly, we extrapolated the downward collision rates for $\rm HCO^+$ 
(Flower 1999) to use for $\rm H^{13}CO^+$ and $\rm DCO^+$.
Upward rates were calculated for detailed balance for each isotope.
The size of each core is fixed at 0.15 pc, and microturbulent velocity is 
assumed constant through each modeled core.

Once the MC code calculates each level population, we simulate specific
molecular line profiles using the virtual telescope (VT) simulation to compare
the modeled profiles with observed molecular line profiles.
We can simulate a molecule with multiple transitions and multiple positions 
simultaneously in a specific core with the parameters such as beam sizes 
and main beam efficiencies that we used in the actual observations.
The remaining lack of realism is caused by geometric 
differences between a 1-dimensional simulation and the actual core.  
As seen in Figure 2, the three cores are not perfectly spherically symmetric,
but instead they are elongated. In order to compare simulated line profiles with
real observations, we averaged the lines that have been observed at the same
distance from the centers, e.g., the lines in the off-positions of 
$\pm30\arcsec$ in $\alpha$ and $\delta$. 
However, some lines have not been averaged in all four 
directions because the observation did not cover all parts. 
We need to map cores fully and use a 2-dimensional code to do more accurate 
comparisons.  
 
We model the transitions of four molecules, \dc18o ($\rm J=2-1$ and 
$\rm J=3-2$), 
$\rm DCO^+$ ($\rm J=3-2$), $\rm H^{13}CO^+$ ($\rm J=1-0$ and $\rm J=3-2$), and 
$\rm HCO^+$ ($\rm J=3-2$).
If the gas and dust of a core are well mixed and well coupled, we can apply 
the distributions of density and temperature of the dust to the gas. 
However, for a density lower than about $\rm 10^4~cm^{-3}$, the gas temperature 
is not coupled well with the dust temperature. 
The excitation temperatures of the lines that have high
critical density such as $\rm DCO^+$ and $\rm H^{13}CO^+$ are mainly dependent
on density so that the inner, denser region mainly contributes to the 
intensities of those molecular lines, and the outer, less dense region does not 
change our results much. 
However, we should in principle correct for the 
difference between gas and dust temperatures in the outer, less dense parts 
of cores, especially for the molecular lines such as \dc18o 
lines that have low critical density.
We leave this correction for a future work.

\subsection{Physical Models}

We use two different types of physical models: Bonnor-Ebert spheres
(Bonnor 1956, Ebert 1955);
and Plummer-like models (Whitworth \& Ward-Thompson 2001).
Both these models allow an initial density distribution that is
nearly constant for small radii, but approaches a power law
at large radii. The Bonnor-Ebert sphere provides a good
fit to the dust continuum emission of the three PPCs that we are modeling
(Evans et al. 2001). The Plummer-like model can fit the dust continuum
data for L1544 (Whitworth \& Ward-Thompson 2001). It has the advantage
that it provides a simple solution for the density and velocity field
during a free-fall collapse from the initial state, thereby providing
a velocity field, which we need for modeling some of our lines. 
The initial Plummer-like radial density profile 
(Whitworth and Ward-Thompson 2001) is

\begin{equation}
\rho(r,t=0)=\rho_{flat}\left[\frac{R_{flat}}{(R^2_{flat}+r^2)^{1/2}}\right]^\eta.
\end{equation}

Even though this model is very simple, in that it assumes a pressure-free 
collapse, it may be able to explain the later phase of the evolution of PPCs.
Li et al. (2002) show that the evolution of a core driven by 
ambipolar diffusion becomes closer to the evolution of a non-magnetic, 
free-falling core at higher density.

We use the Bonnor-Ebert sphere model
that best fit the dust emission (Evans et al. 2001), and we assume a
purely microturbulent velocity field, 
except for the $\rm HCO^+$ $\rm J=3-2$ lines in L1544 and L1689B and the 
$\rm H^{13}CO^+$ lines in L1544. 
Because those line profiles show deep self-absorption features indicative of 
systematic radial velocities, 
we use a Plummer-like model for them. To constrain the various
parameters in the Plummer-like model, we modeled the dust emission,
as was done by Evans et al. (2001). The resulting values that fit best
the data for both L1689B and L1544 are as follows:  $\eta=3$, the initial
$\rho_{flat}$ (t=0) of $\rm 5\times 10^5~cm^{-3}$, and $ R_{flat}$ of
2800 AU.  Whitworth and Ward-Thompson found a steeper density structure 
($\eta=4$) and a bigger $R_{flat}$ (5350 AU) because they used an 
isothermal sphere for their model. When the temperature decline toward
the center is included, a smaller $R_{flat}$ is needed.
The evolutionary timescale that fits the dust emission well is half the
timescale at which the central point-mass is formed (the start of the 
Class 0 stage). 
The maximum velocity in this model is about 0.08 km/s, which is consistent with
the results of Tafalla et al. (1998) and Williams et al. (1999) in L1544.
The resulting density structure is compared to that of the best fitting
Bonnor-Ebert sphere in Figure 4a; they are very similar. Likewise,
the resulting dust temperature structure is very similar (Figure 4b).
The velocity field from the Plummer-like model is shown in Figure 4c.

While the density and temperature structure of the best-fitting Plummer-like 
model is very similar to that of the best fitting Bonnor-Ebert sphere, 
the former does contain a systematic velocity field. To check whether that 
field affected the results for lines modeled with the Bonnor-Ebert sphere, 
we modeled all other lines using the Plummer-like model.
The velocity structure does not affect the line profiles of \dc18o, 
$\rm DCO^+$, and $\rm H^{13}CO^+$; the two models have given the same results.

\subsection{Results of Models}

The main purpose of this modeling is to find how much a specific molecule
is depleted, and how the depletion is distributed with radius by simulating 
molecular line profiles at multiple positions within 
a core and by comparing the simulated line profiles with observed line profiles.
We tried several functional forms, such as exponential, power laws, 
Plummer-like, and step functions, for the distribution of a molecular 
abundance. 
In addition, we applied the results of Li et al.'s dynamical and chemical 
model (2002) to our modeling, but their model has too high a systematic velocity 
to fit the observed molecular line profiles.
However, the group is improving their chemical and dynamical models while 
communicating with us.
Figure 5 and Table 6 shows the comparison between results of different 
functional forms of depletion with radius in the C$^{18}$O 2$-$1 line 
simulations.
We used integrated temperatures to calculate $\chi ^2$ in each model.
Among those distributions, a step functional distribution fits best 
the observed line profiles with radius. This seems reasonable because the
results of chemical models show a sharp decrease in abundance with density
or time. 
Here, we mainly show the results of the modeling with the abundance 
distribution of a step function that has three free parameters: 
the undepleted abundance of a molecule ($X_0$), the fractional depletion 
($f_D=X/X_0$), and the radius inside which a molecule is depleted 
($r_D$). Each can be different in different species. 
Therefore, the total depletion of a molecule is dependent on the combination of
$f_D$ and $r_D$. We emphasize that the derived depletion factor from this 
modeling has an uncertainty of about a factor of 3 because of the uncertainty of
the density structure calculated from the dust analysis. 

We compared the distributions of observed and modeled integrated temperatures 
with radius to find the best fits. Figure 6.a shows the distribution of the 
reduced $\chi ^2$ of the models in the C$^{18}$O 2$-$1 and 3$-$2 lines of 
L1512 in the space of $f_D$ and $r_D$. 
The depletion radius is well constrained around 0.075 pc, but the fractional 
depletion is constrained only to be greater than about 25. 
This trend occurs in every molecule that is significantly depleted, 
so we consider $f_D$ to be the lower limit of the depletion factor.
In other words, the data are consistent with complete depletion inside some
$r_D$.
However, the distribution of $\chi ^2$ constrains only the upper limit of 
the depletion radius in H$^{13}$CO$^+$ and DCO$^+$ of L1689B and 
in DCO$^+$ of L1544, which show small $r_D$ and $f_D$ (see Table 8 and Table 9).
For example, the $\chi ^2$ in H$^{13}$CO$^+$ of L1689B changes within a factor 
of about 2 (from 3 to 7) in all range of $f_D$ within $r_D < 0.011$ 
(Figure 6.b).  
As we mentioned above, the density structure has uncertainty of 
a factor of about 3, so the calculated depletion in this model could have the 
same uncertainty.

\subsubsection{$C^{18}O$ $J=2-1$ and $J=3-2$}

The best fits of \dc18o $\rm J=2-1$ and $\rm J=3-2$ line profiles in the three 
cores are shown in Figure 7. The parameters of the best fit models are 
summarized in Table 7. 
We used the standard \dc18o abundance of $4.82\times10^{-7}$ for 
the undepleted abundance ($X_0$) for all cores, which was used to
calculate the column density of $\rm H_2$ ({$\S$ 4.2.1}).
According to the results, CO molecules are depleted within 0.075 pc 
(about 110\arcsec) by a factor of 25 in L1512, so CO is depleted everywhere
in our C$^{18}$O map. 
In the case of L1544, the depletion factor is 25 within 0.045 pc 
(about 70\arcsec). \dc18o $\rm J=2-1$ and $\rm J=3-2$ line profiles simulated 
simultaneously at different positions from the center fit the observed line 
profiles well in L1512 and L1544.
On the other hand, we could not find a model to fit well all the observed line 
profiles from the center to the off-position of $90\arcsec$ in L1689B.  
The antenna temperatures of observed $\rm J=2-1$ and $\rm J=3-2$ lines are 
almost the same or increase outward from the center in L1689B, 
even though the outermost lines are not well sampled. 
This might indicate that the large molecular cloud surrounding this source 
(Loren 1989) contributes to the observed \dc18o lines.
We tested a core that has a warm envelope surrounding an inner, denser core.
The core has the same density and temperature structures inside 0.15 pc 
as those calculated from dust emission, and a constant density and temperature 
of $10^3$ cm$^{-3}$ and 50 K from 0.15 to 0.6 pc, which are based on
the $^{13}$CO observations (Loren 1989). The temperature of 50 K is the upper
limit in the $\rho$ Oph molecular complex.
The modeled line profiles are shown in Figure 7c, which fit better the observed
line profiles than those in the previous model that does not have a warm 
envelope.  
In this model, the abundance distribution is the same as the previous one.
In order to avoid confusion by the envelope, we need to use a more rare 
isotope or higher J transition to trace the high density core.
Even though L1512 and L1544 are believed to be well isolated, they might also
be surrounded by bigger clouds.
The modeled results, however, fit well the observed \dc18o $\rm J=2-1$ to the 
off-position of 90\arcsec. The influence of bigger structures in L1512 and
L1544 does not seem significant compared to L1689B.  

Jessop and Ward-Thompson (2001) modeled the C$^{18}$O J$=$2$-$1 and J$=$3$-$2 
lines of L1689B using a density structure with the same central density 
($\rho_{flat}$) and the same $R_{flat}$ as calculated by Andr\'e et al. 
(1996)  and a constant temperature. After comparing to 
observations they argued that the flat distribution of C$^{18}$O is caused by 
the significant (about 95\%) depletion of CO. 
They pointed out that the gas temperature must be above 14 K
to have better agreement with observations. However, they compared the model to
the observation within 32$\arcsec$, which cannot give a good constraint 
on the outer envelope. 
In addition, the lower density and higher temperature at the inner region 
than those assumed in this study make the transitions optically thinner than 
the actual lines, so the depletion factor to fit the observations could be 
overestimated.
We modeled \dc18o lines using our density structure and a constant temperature 
of 14 K.
The best-fit model has $r_D \sim 0.035$ and $f_D \sim 10$, 
which are bigger than our best-fit model, where the 
temperature structure is calculated from the analysis of dust emission.
The modeled \dc18o J$=$2$-$1 lines within 60$\arcsec$ are stronger than the 
observed lines. However, the line at 90$\arcsec$ is still weaker than
the observed line even though it is stronger than that in our best-fit model
without a warm envelope. 

\subsubsection{$H^{13}CO^+$ $J=1-0$ and $J=3-2$}

Except for the C$^{18}$O lines, $X_0$ is not fixed in modeling molecular 
lines, but it can be easily decided by the outermost line profile. 
The results of the best models in $\rm H^{13}CO^+$ $\rm J=1-0$ and $\rm J=3-2$ 
are shown in Figure 8 and Table 8. We have observed the $\rm H^{13}CO^+$ 
$\rm J=1-0$ line with a $20\arcsec$ grid out to the distance of $60\arcsec$ 
from the center with an $18\arcsec$ beam.
The $\rm J=3-2$ line has been observed with $30\arcsec$ spacing with a 
$26\arcsec$ beam. 
In L1512, the $\rm H^{13}CO^+$ $\rm J=3-2$ line, whose critical density is about
$10^6$ cm$^{-3}$, was not detected, so we simulated
only the $\rm J=1-0$ line. The depletion factor in L1512 is 25 within 0.021 pc 
(38\arcsec). 
In L1689B, the best-fit model of $\rm J=1-0$ line also fits $\rm J=3-2$ 
line very well. The depletion factor and radius in L1689B are much smaller 
than those in L1512. However, the undepleted abundance of $\rm H^{13}CO^+$ 
in L1512 is 1.4 times bigger than in L1689B.  

The observed $\rm H^{13}CO^+$ $\rm J=1-0$ line, whose critical density is about
$10^5$ cm$^{-3}$, shows a blue skewed profile in L1544. 
Contracting cores show stronger blue peaks and 
self-absorption dips in optically thick lines (Myers et al. 1996).
Evidence of inward motion in L1544 has been observed by Tafalla et al. (1998) 
and Williams et al. (1999).   
Therefore, we used the best-fit Plummer-like model in the simulation of the
$\rm H^{13}CO^+$ $\rm J=1-0$ line.
In L1544, models fit well the shapes of the observed $\rm H^{13}CO^+$ 
$\rm J=1-0$ line profiles, but the modeled $\rm J=3-2$ line is {\it stronger}
than the observed line by a factor of 2. The rms noise of this line is
about 0.015 K (5$\sigma$ detection), and the baseline is poor, so the 
discrepancy may not be very significant.
There are several possibilities.
If there is molecular gas surrounding L1544, this less dense gas would increase 
the $\rm J=1-0$ line more than the $\rm J=3-2$ line. 
Therefore, our models would 
overestimate the abundance of the inner, denser core to fit the 
$\rm J=1-0$ line. As a result, the modeled $\rm J=3-2$ line can be stronger 
than the observed line.
Another possible explanation is that the depletion fraction increases toward 
the denser, inner region. Since the $\rm J=3-2$ line traces denser gas than 
the $\rm J=1-0$ line, the $\rm J=3-2$ line in our models where we used a 
step function of the abundance would be stronger than the observed line 
if $\rm H^{13}CO^+$ is depleted more at the center.
The uncertainties in the observed line or in the density and temperature 
structures could explain the difference between the modeled and observed 
lines.

\subsubsection{$DCO^+$ $J=3-2$}

The modeling of $\rm DCO^+$ $\rm J=3-2$ does not constrain free parameters well 
because the observations in the line do not cover positions far from the center.
The results are shown in Figure 9 and in Table 9. 
In L1512, only the center position was observed, so we just used a constant
abundance to fit the line profile. The abundance of $\rm DCO^+$ that fits the 
line profile well is $2.8\times10^{-10}$.
However, the line profiles of two positions in L1544 and three positions 
in L1689B were compared with the results of models.
The factor and the radius of the depletion of $\rm DCO^+$ molecule are much 
smaller than those of CO molecules calculated by simulating 
\dc18o $\rm J=2-1$ and $\rm J=3-2$ lines. 
This result is consistent with the results of the chemical model of Caselli 
et al. (2002). 
The radius of the depletion is less (20\arcsec) than the beam size (35\arcsec) 
in L1689B and similar to the beam size in L1544.

\subsubsection{$HCO^+$ $J=3-2$}
 
Infall asymmetry in PPCs has been observed (Tafalla et al. 1998, Lee et al. 
1999, Gregersen and Evans 2000) in optically thick lines.
If the gas of the outermost envelope of a core that has inward motion 
has sufficient optical depth and has a lower excitation temperature than the 
gas of the inner envelope in a given molecular line, the molecular line shows
infall asymmetry, in which the intensity ratio of blue peak to red peak 
increases with time (Gregersen et al. 1997).
We model one of the optically thick lines, $\rm HCO^+$ $\rm J=3-2$ in the 
three PPCs and compare the results with the observational results of Gregersen 
and Evans (2000).
In order to model an $\rm HCO^+$ $\rm J=3-2$ line that has infall asymmetry, 
we used the best-fit Plummer-like model.
Figure 10 shows the difference in MC modeling of the $\rm HCO^+$ $\rm J=3-2$ 
line for two different physical models: Bonnor-Ebert sphere and 
Plummer-like model. The latter includes an infall velocity structure and
produces a blue asymmetry. 
The $\rm HCO^+$ $\rm J=3-2$ line in L1512 does not show a self-absorption dip 
and asymmetric profile so that we model the 
line in L1512 using the best-fit Bonnor-Ebert sphere.

The results of modeling $\rm HCO^+$ $\rm J=3-2$ in the three PPCs are shown in 
Figure 11, and in Table 10.
Because L1512 has only one spectrum at the position 30$\arcsec$ away from 
the center, we used a constant abundance of $\rm HCO^+$ for this core.
The abundance of the best fit is about $1\times10^{-9}$.  
In L1544, the intensity of the $\rm HCO^+$ $\rm J=3-2$ line at 30$\arcsec$ is 
half that at the center, unlike the $\rm H^{13}CO^+$ 1-0 line, 
whose intensity is almost flat to 40$\arcsec$.
As a result, an exponential abundance distribution as well as a step function 
also fits well the line profiles. 
In L1689B, the $\rm HCO^+$ $\rm J=3-2$ line has a stronger red-peak asymmetry 
at the center
and 15$\arcsec$ away, but it has a stronger blue-peak asymmetry at 30$\arcsec$ 
and 45$\arcsec$. This complication of line profiles causes difficulty in 
fitting the exact line profiles in this core. Therefore, we tried to fit 
the integrated temperatures of the lines with radius. 
The exponential function of the abundance shows 
similar results to the step function. In order to test the functional 
form of the abundance, we need a more extended map.
The interesting result is that the $\rm HCO^+$ 
$\rm J=3-2$ line of L1689B has a similar strength to that of L1544, unlike 
$\rm H^{13}CO^+$ $\rm 1-0$, which is stronger 
in L1689B than in L1544. 
This might be due to the fact that L1689B is embedded in a larger molecular 
cloud that causes more absorption of the $\rm HCO^+$ 3$-$2 line.

On the whole, the abundance of $\rm HCO^+$ in this result is very small 
compared to the abundance that is expected from $\rm H^{13}CO^+$ and 
the typical ratio of C and $\rm ^{13}C$ in the local interstellar medium of 
about 77 (Wilson and Rood 1994). However, our results 
show the abundance ratio of $\rm HCO^+$ and $\rm H^{13}CO^+$ is approximately from 3 to 
15 in the three PPCs.   
(We do not think that the ratio in L1512 is correct because we used only one 
spectrum in each line to calculate the abundance.)
In addition, the depletion factors and the depletion radii calculated from
the $\rm HCO^+$ line are different from those calculated from the 
$\rm H^{13}CO^+$ lines. 
These differences and the low abundance ratio may result from 
the gas components outside the cores that we modeled in this study 
because the $\rm HCO^+$ line can trace less dense 
gas than $\rm H^{13}CO^+$. As a result, the $\rm HCO^+$ line can be absorbed by 
the surrounding material significantly, but the $\rm H^{13}CO^+$ line may be not 
absorbed as much as the $\rm HCO^+$ line.
In our models, we limited the size of every core to 0.15 pc so that our models 
cannot account for the gas outside the core. 
Then, we would have underestimated the abundance trying to fit a line that 
has been weakened by the absorption by the surrounding gas component.  
However, this explanation is just one possibility. We cannot ignore the 
uncertainties of many other parameters, so we leave this as future work. 

\section{DISCUSSION}

\subsection{Depletion}

We concluded that CO is significantly 
depleted in L1512 and L1544. It is difficult to estimate the 
depletion of CO in L1689B because of the gas surrounding this core 
and the optical depths of the lines used, but the depletion appears to be
much less.
The $\rm HCO^+$ molecule is depleted in L1512 and L1544, but we do not see 
significant depletion of this molecule in L1689B, based on modeling
the $\rm H^{13}CO^+$ $\rm J=1-0$ line. 
The optical depth and possible surrounding material also affect 
the calculation of the depletion of $\rm HCO^+$ in L1544.
$\rm DCO^+$ is not significantly depleted in any of the PPCs. 
This result is consistent with results in many other starless cores where the 
distribution of DCO$^+$ fits well the distribution of dust continuum 
(Myers 2002, personal communication). 
However, we need to use better resolution to do a more precise study because 
the calculated depletion radius is not bigger than the actual beam size 
(35$\arcsec$). Caselli et al. (2002) show that the depletion
radius of DCO$^+$ is about 3000 AU in L1544.   

The analysis of the integrated intensity shows that the earliest-time molecule 
CCS is not peaked toward the centers of any of the PPCs, unlike the dust 
emission.
Thus, CCS is depleted significantly in all three.
As we saw, however, the CCS intensity in L1544 is twice stronger than that 
in L1689B even though the dust emission of L1689B and L1544 is similar.
This result suggests that the depletion of CCS is connected to the 
depletion of CO.
Li et al. (2002) modeled chemical abundance changes in PPCs using a
dynamical model with a magnetic field, which drives the ambipolar diffusion, and
they got a relatively high abundance of CCS.
They indicated that this abundance could result from the interplay between 
significant depletion of CO molecules and late-time hydrocarbon chemistry.
Ruffle et al. (1997, 1999) also showed the same result in their
models of the time-dependence of several molecules.
According to their models, the abundance of CCH and CCS
have later secondary maxima only when the freeze-out timescale is long 
compared to the chemical timescale or the collapse timescale.
CCS increases quickly to the first maximum of its abundance in time and is 
depleted very early as the density increases because CCS is destroyed by 
increasing HCO$^+$ or H$_3$O$^+$, and this polar molecule can be tightly bound 
in grain mantles. 
However, in the region where CO molecules are depleted significantly, the 
$\rm C^+$ ion reacts more with molecules that are not oxygen-bearing species 
so that the precursors of CCS such as CCH increase again. 
This results in a second increase of the CCS abundance.
The trend has not shown up in the models of Bergin \& Langer (1997) 
and Aikawa et al. (2001) because the models used dense
initial conditions in which a core evolves very quickly dynamically, 
and CCS did not show the second maxima.
We think that the CCS molecule is possibly in the second 
enhancement in the late stage of chemical evolution resulting from the 
significant depletion of CO in L1512 and L1544. 
The integrated intensity of the CCS line shows bumps in the center regions of 
L1512 and L1544 (Figure 3). If CCS is depleted with time without the second 
enhancement, we should see holes rather than bumps. 
The other possible explanation of the enhancement of CCS is mixing of some
atomic carbon from a translucent envelope into the core (van Dishoeck 2002, 
personal communication).

According to the comparison between the intensity of $\rm N_2H^+$ lines and
the dust emission intensity, this molecule is possibly depleted in L1512, 
but not in L1544 and L1689B, where its intensity follows well the dust emission 
intensity. $\rm N_2H^+$ is known as a very late-time molecule, but Bergin et al
(2002) have shown that $\rm N_2H^+$ is depleted in a cold dark cloud, B68.
The weak intensity of $\rm N_2H^+$ line in L1689B, compared to in L1544,
might indicate that this molecule has not yet reached its maximum because of
less abundant precursors (N$_2$ and H$_3^+$) but a more abundant destroyer (CO).

Bergin and Langer (1997) modeled chemical evolution in PPCs including gas-grain
interactions as well as gas-phase reactions and compared two kinds of grain
mantle properties: a weakly bound CO mantle  and tightly bound $\rm H_2O$
mantle. Their results show that sulfur-bearing molecules such as CS and SO
are very sensitive to density and depleted seriously as the density
of a core increases, regardless of the properties of dust grains.
On the other hand, CO and $\rm HCO^+$ molecules are depleted in the model
with $\rm H_2O$ grain mantles but remain in gas phase in the model with CO
grain mantles. In contrast, $\rm NH_3$ and $\rm N_2H^+$ do not show
depletion in any model because of the low binding energy of $\rm N_2$ to
$\rm H_2O$ and CO mantles. 
Therefore, the comparison of the results of their chemical models to 
our results indicates that the dust grains in the three PPCs have tightly
bound $\rm H_2O$ mantles.
In addition, Aikawa et al. (2001) predicted the abundance distributions of 
molecules by numerical chemical models simply by applying a constant delay 
factor to a Larson-Penston model and considering different sets of adsorption 
energies, and they compared their results with the observed molecular 
abundances in L1544.
According to their results, CCS and CO molecules are depleted,
but $\rm N_2H^+$ and $\rm NH_3$ molecules are more abundant in more slowly 
collapsing cores at the density peak. 

To summarize our results on depletion, CCS, CO, and $\rm HCO^+$  
are significantly depleted in L1512 and in L1544. On the other hand, the
depletion of $\rm DCO^+$ in the two cores is not significant.
$\rm N_2H^+$ is not depleted in L1544, but it is possibly depleted in L1512.
In contrast, only CCS is significantly depleted in L1689B.
It is hard for us to estimate the depletion of CO in L1689B 
because of the optical depth effect and the material surrounding the cores.
 
\subsection{Ionization}

The coupling between ions and the magnetic field is crucial in regulating star 
formation rates if cores are magnetically subcritical.  
The timescale of ambipolar diffusion depends on the ionization fraction
($\rm t_{AD}\approx 2.5\times 10^{13} \it x(e)$, Shu et al. 1987). 
In addition, the gas-phase chemical reactions are also dependent on the 
ionization fraction.
Therefore, calculating the ionization fraction accurately is significant
for understanding the chemistry as well as the timescale for ambipolar diffusion. 
However, it is not easy to calculate the ionization fraction in protostellar
cores because it depends on the depletions of molecules as well as density
structures and cosmic-ray ionization rate.  

We used equation (16) in Caselli (2002) to calculate the 
ionization fractions:

\begin{equation}
x(e)=\frac{2.7\times10^{-8}}{\rm [DCO^+]/[HCO^+]}-1.9\times10^{-7}\left[\frac
{1}{\rm f_D}+\frac{\rm x(O)/f^{'}_D}{10^{-4}}\right],
\end{equation}

\noindent where $\rm f_D$ is the depletion factor of CO, x(O) is the
abundance of atomic oxygen ($1.5\times10^{-4}$, Caselli et al. 1998), 
$\rm f^{'}_D$ is the depletion factor of O, 
and $\rm [DCO^+]/[HCO^+]$ is 
the abundance ratio of two molecules, which can be calculated by comparing 
the column densities of $\rm DCO^+$ and
$\rm H^{13}CO^+$ and using 77 $(\pm 7)$ for the $\rm ^{12}C/^{13}C$ ratio
(Wilson \& Rood 1994).
Caselli et al. (1998) and Caselli (2002) derived this equation assuming 
that (i) the main source of $\rm HCO^+$ is the reaction between
$\rm H_3 ^+$ and CO, (ii) the deuterium fractionation is due to the reaction
between $\rm H_3 ^+$ and HD, and (iii) molecular ions are destroyed mainly
by electrons and neutral species such as CO and O.
In fact, Caselli et al. (2002) have showed that N$_2$H$^+$ and N$_2$D$^+$ 
are better tracers of ionization fraction because HCO$^+$ and DCO$^+$ are 
possibly depleted in the central region.
However, we do not have N$_2$D$^+$ data, so we use $\rm DCO^+$ and 
$\rm HCO^+$, which can give correct information in the region with 
$r>3000$ AU.  

We calculate, for each line of sight, the average $x(e)$ along the line of
sight. This approach is not completely self-consistent because the method is
only valid for homogeneous clouds. We use our model to simulate observation 
with the same beam size.
The column density of each molecule has been
calculated from the simulated line based on the assumption of an optical thin
line.  
Figure 12 shows the ionization fraction with radius in the three PPCs.

There are some caveats in this calculation. First, we could not assess the 
depletion factor of CO in L1689B correctly because of the optical depth 
problem and the possible material surrounding the core ($\S$ 5.2.1). 
Second, the model of $\rm H^{13}CO^+$ lines in L1544 does not match the 
observations very well so that the calculated column density from the simulated 
$\rm H^{13}CO^+$ $\rm J=3-2$ has more uncertainty. Third, since we do not
have enough observed data to assess the depletion radius of $\rm DCO^+$
in L1512, the column density of $\rm DCO^+$ is uncertain in 
this core too. In addition to the caveats related to models, the equation that
we used to calculate ionization fraction includes the abundance and the 
depletion fraction of O that are not well known. 
We supposed that the depletion of atomic oxygen might be proportional to the 
depletion factors of other molecules.
We used $\rm f^{'}_D$ of 3, 2, and 1 for L1512, L1544, and L1689B, 
respectively.

In spite of these caveats, the three cores have similar average ionization 
fractions (about $5\times10^{-7}$) along the line of sight toward the centers.
These results are similar to the ionization fractions calculated in other
protostellar cores (Caselli et al. 1998) but much bigger than the result
(about $10^{-9}$ at the center) of Caselli et al. (2002). 
Caselli et al. (1998) have used the integrated temperatures of 
the actual line profiles, but Caselli et al. (2002) have used the results of
their ``L1544 best-fit" chemical model, which takes into account the cloud
density structure, molecular freeze out, and the recombination of molecular
ions on grain surfaces, in order to infer the electron fraction as a function
of cloud radius.
Our data do not rule out still smaller values of $x(e)$ at small radii.
The ionization fraction in L1512 and L1544 decreases toward the center, which 
trend has been shown in Caselli et al. (2002) and Caselli (2002). 
In contrast, in L1689B, it increases toward the center. 
As a result, the ionization fraction of L1689B around 0.06 pc is smaller
than those of L1512 and L1544 by a factor of 10. 

\subsection{Timescales}

The timescale of ambipolar diffusion ($\rm t_{AD}\approx 2.5\times 10^{13} 
\it x(e)$, Shu et al. 1987) calculated from the above ionization fraction 
toward the centers is about $1.25\times10^7$ years. This timescale applies to 
cores that are magnetically subcritical.
This ambipolar timescale represents the duration of contraction from the 
present stage to the stage of Class 0.   
These are much bigger than the free fall timescale ($\approx 10^6$ years).
However, $x(e)$ could be much smaller at the center.
Even though these three cores have similar ambipolar diffusion timescale 
toward the centers, our results suggest that they have different chemical 
status from each other.

L1512 seems to have evolved chemically more than L1544 showing that CCS, CO, 
HCO$^+$, and possiblely N$_2$H$^+$ are depleted. 
Caselli (2002) modeled a core that is less dense and less centrally
concentrated than L1544 to show the chemistry in an early stage in the
dynamical evolution. According to their result, molecules in the model core
are not significantly depleted. The model core is similar to L1512 in density
structure, but surprisingly, L1512 shows a bigger depletion radius of CO than
L1544. In addition, N$_2$H$^+$ at the center of L1512 might start to be depleted.
This result could result from the slower contraction of L1512 than L1544 
so that L1512 could have a longer time to evolve chemically.   
Another possibility is a more efficient molecular freeze out in 
L1512 because the dust grains in this object are smaller than in the other two 
cores, so that the depletion timescale in L1512 is shorter (van Dishoeck et al. 
1993; Caselli et al., in prep.).  This could be possible because L1512 is less 
dense than the other two cores and has not have enough time for grain coagulation
(Evans et al. 2001).

On the other hand, L1689B has not evolved much chemically even though 
the density structure is very similar to L1544 so that it is in a later stage
of dynamical evolution than is L1512. 
This tells us that the chemical evolution is not a simple function of density,
unlike the results of other chemical models. 
According to this result, we should not ignore the possibility that L1689B could
be magnetically supercritical, and the core has evolved faster than the
other two sources dynamically so that L1689B has not had enough time to evolve
chemically. The broader lines in L1689B (Figure 1 and $\S$ 3.1) 
indicate more active kinematics than the other two sources.  
The total masses of L1544 and L1689B are similar according to the dust modeling.
Therefore, L1689B has to have a weaker magnetic field or a smaller ionization
fraction in order to be magnetically supercritical. 
Crutcher \& Troland (2000) measured the magnetic field along
the line of sight in L1544 ($B_{\rm LOS}\approx11$ $\mu$G) by using the Zeeman 
effect of the OH lines. There is no specific measurement of magnetic 
fields in L1689B. However, Troland et al. (1996) measured a OH Zeeman effect
in the $\rho$ Oph cloud indicating $B<10$ $\mu$G.
In addition, our calculation shows the ionization fraction in L1689B is smaller
than in the other two cores in the outer part, where the calculation is more
credible.
The other possibility of the faster dynamical evolution of L1689B than L1544 is
a larger external pressure (Galli et al. 2002) because L1689B is in the $\rho$  
Oph complex (see \S 5.3.1).

Based on our results, we suggest that the stage of 
dynamical evolution of a core cannot be simply probed by its chemical status 
because the chemical evolution depends on the size of dust grains or 
the absolute timescale, during which the core has been in a given environment, 
as well as its density structure that shows the relative stage of dynamical 
evolution.
We summarize the dynamical and chemical evolutionary stages of each core
in Table 11.

\section{SUMMARY}

The summary of our results is as follows:

\noindent 1) The difference between the column densities inferred from dust 
emission and the emission of a given molecular line can partially result from 
the optical depth of the line. However, the 
depletion of a molecule must explain the remaining difference. 
In L1689B, the main cause of the difference between the dust emission and 
the CO emission is the optical depth in the lines.
However, in L1512, the depletion of CO is the major source of the difference 
between the dust emission and the CO emission. 

\noindent 2) In our Monte Carlo simulation of molecular lines, a step function 
with depletion in the inner region produces the best fit to the abundance 
variation of a molecule in PPCs. 

\noindent 3) The CO and $\rm HCO^+$ molecules are depleted significantly in 
L1512 and L1544, but not in L1689B. The gas around L1689B makes it difficult
to calculate the actual depletion of CO.
 
\noindent 4) The depletion of $\rm DCO^+$ is not substantial in the three PPCs.

\noindent 5) The CCS molecule is substantially depleted in the three PPCs.
However, CCS might be in the second enhancement in L1512 and L1544 because of
the significant depletion of CO. 

\noindent 6) The distribution of the integrated intensity of $\rm N_2H^+$ $\rm 
J=1-0$ suggests the possible depletion of $\rm N_2H^+$ in L1512.
On the other hand, $\rm N_2H^+$ follows dust emission very well in L1544.  
In L1689B, $\rm N_2H^+$ has not had enough time to reach its maximum.

\noindent 7) The velocity structures calculated by a simple dynamical model that 
uses a Plummer-like density profile and a free-fall collapse model can produce 
the asymmetry of $\rm HCO^+$ $\rm J=3-2$ lines in L1544 and L1689B. 
However, the abundance calculated by modeling the lines is much smaller than
what is expected from modeling $\rm H^{13}CO^+$.
The surrounding gas could possibly account for the low $\rm HCO^+$ abundance
of the models.

\noindent 8) The ionization fraction is similar toward the centers of the three 
cores even though L1512 is less centrally condensed than L1544 and L1689B. 
However, L1512 has a bigger depletion radius of CO than L1544 and possible
depletion of N$_2$H$^+$ at the center. 
In contrast, L1689B does not show the maturity in chemical evolution compared
to the other two cores.
This difference suggests that the chemical evolution depends on the size of 
dust grains or the absolute timescale during which a core stays in a given 
environment as well as the density structure of the core that shows the relative 
dynamical evolutionary stage.  

\noindent 9) The evolved density structure and the young chemistry of L1689B 
suggest that this core is possibly experiencing a free-fall collapse rather 
than ambipolar diffusion. 
As a result, L1689B evolves dynamically too fast to evolve chemically. 
This would be possible if the magnetic field in L1689B is weak enough, or the
external pressure is big enough to make L1689B magnetically supercritical.

\section{Acknowledgments}

We are grateful to Paola Caselli, the referee of this paper for helpful 
comments.
We thank the staff of the Caltech Submillimeter Observatory and the Nobeyama
Radio Observatory for assistance in using 10.4 m and 45 m telescopes. 
We also thank E. van Dishoeck, P. Myers, and J. Rawlings for their useful 
comments. We are very grateful to Z.-Y. Li and V.I. Shematovich
for providing the data from their models. 
We thank State of the Texas and the NSF (Grant AST-9988230) for support. 
J.-E. Lee thanks the Rotary Foundation for support through 
an Ambassadorial Scholarship.

\clearpage

%%%%%%%%%%%%%%%%%%%%% Figures %%%%%%%%%%%%%%%%%%%%%%%%%%%%%%%%%%%%%%%%%%%%%%%%
\begin{figure}
\figurenum{1}
\plotone{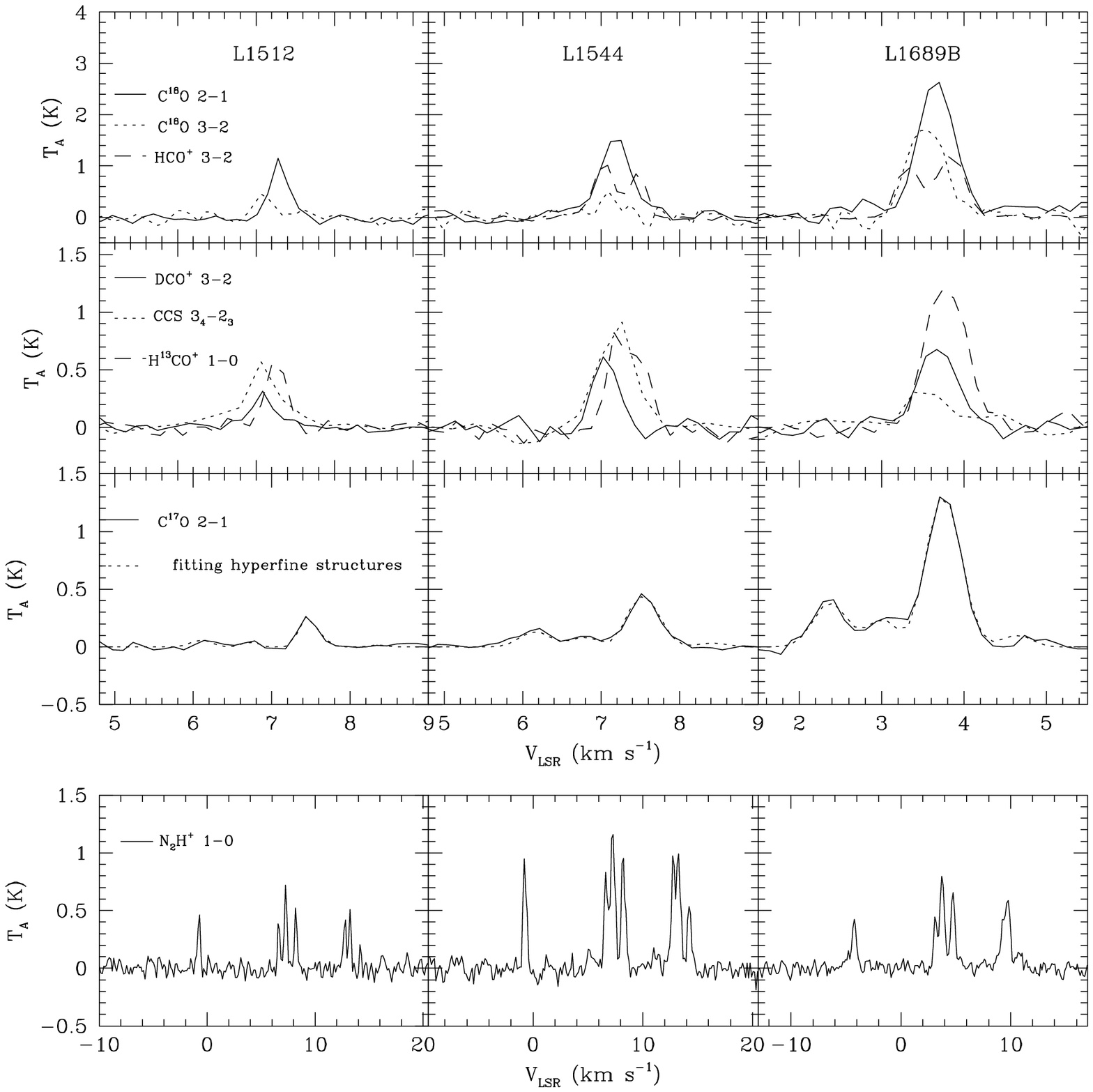}
\figcaption{The observed spectra toward the dust peaks of three PPCs. 
The dotted line in the boxs of the third row shows the results of fitting the 
hyperfine 
structure of $\rm C^{17}O$ $\rm J=2-1$ line profiles using solutions in CLASS. 
}
\end{figure}

\begin{figure}
\figurenum{2.a}
%   \leavevmode
%  \special{psfile=f2a.ps hoffset=-40 voffset=-250 hscale=50
%      vscale=50 }
\plotone{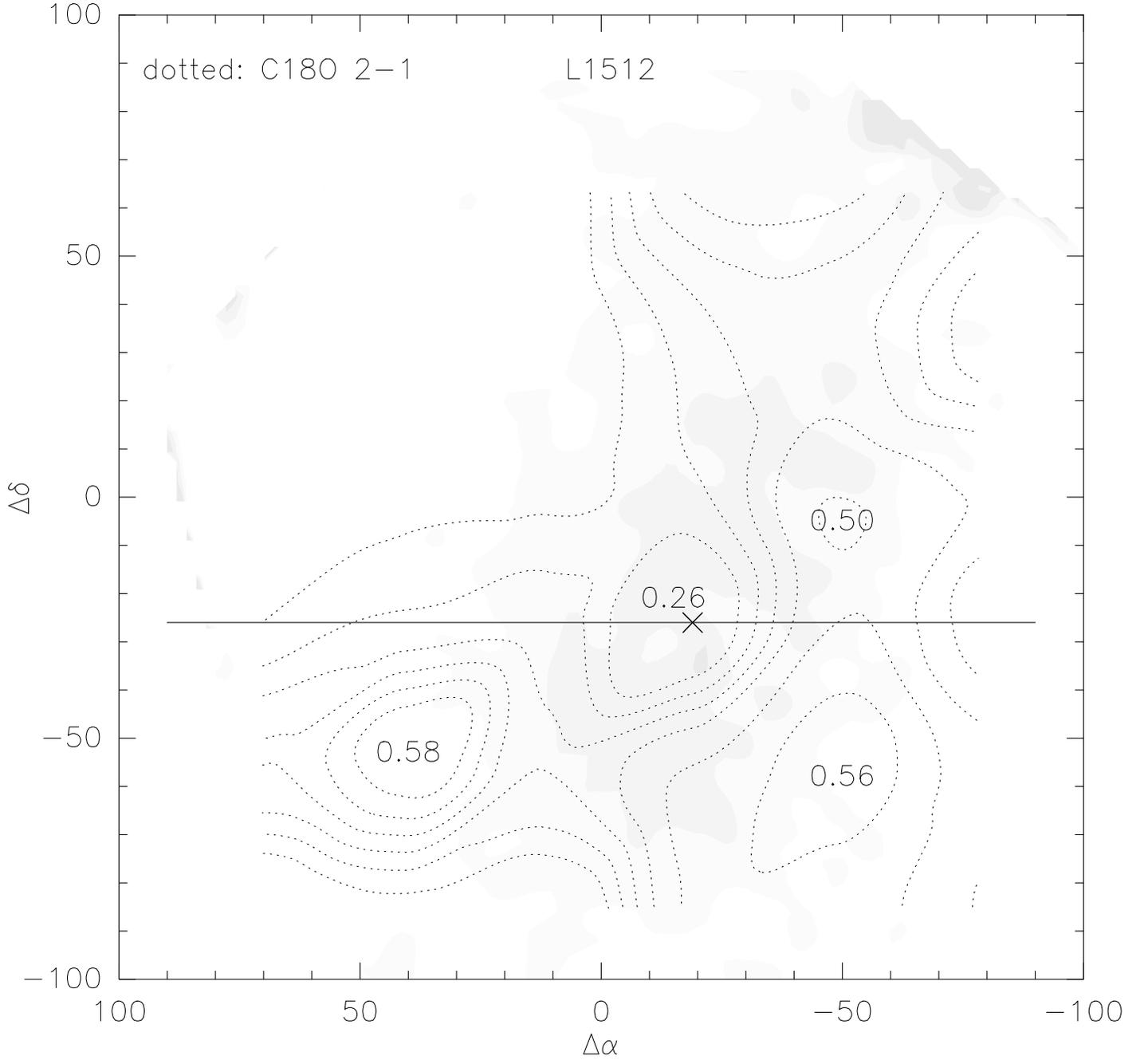}
%  \special{psfile=f2b.ps hoffset=200 voffset=-250 hscale=50
%      vscale=50 }
%  \special{psfile=f2c.ps hoffset=80 voffset=-500 hscale=50
%      vscale=50 }
%\vskip 17cm
 \figcaption{
 These maps show the distributions of 850 $\rm \mu m$ dust continuum
intensity (grey scale, Shirley et al. 2000) and integrated intensities 
of $\rm C^{18}O$ $\rm J=2-1$ line
(dotted contour scale) and $\rm DCO^+$ $\rm J=3-2$ line (solid contour scale).
The (0,0) position is the reference position in Table 1.
The contour levels are from 3 $\sigma$ intensity to peak intensity by
1 $\sigma$ intensity.
For L1512, we put integrated intensity values in several points to
distinct a hole and peaks.
Cross marks indicate the central peak positions of 850 $\rm \mu m$ dust
continuum emission, and horizontal lines show the cuts that we compare
integrated intensities of CCS, $\rm H^{13}CO^+$, and $\rm N_2H^+$ transitions 
or the column densities calculated from $\rm C^{18}O$ $\rm J=2-1$ and 
$\rm C^{17}O$ $\rm J=2-1$ transitions.
 }
\end{figure}

\begin{figure}
\figurenum{2.b}
\plotone{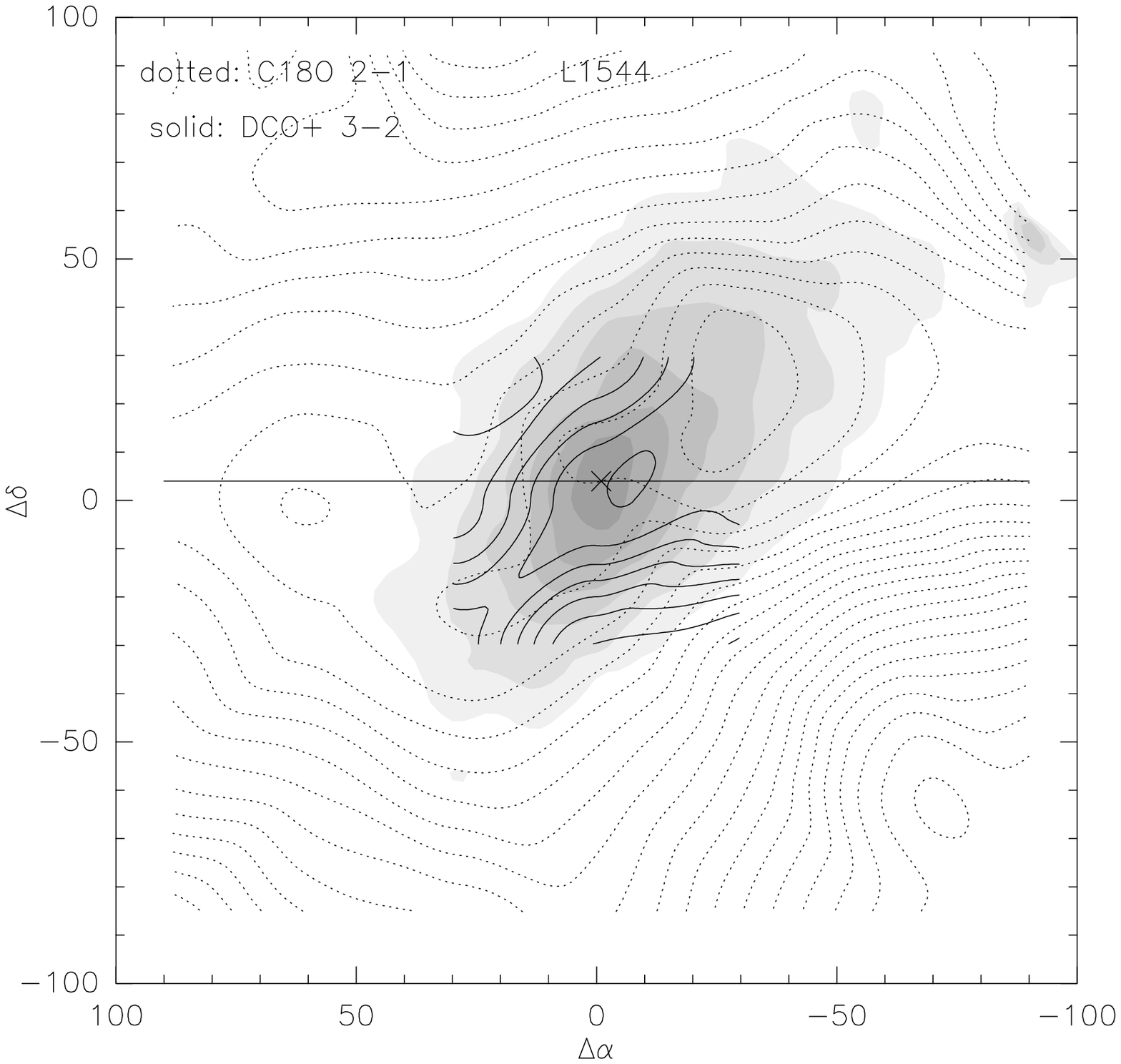}
\figcaption{
}
\end{figure}

\begin{figure}
\figurenum{2.c}
\plotone{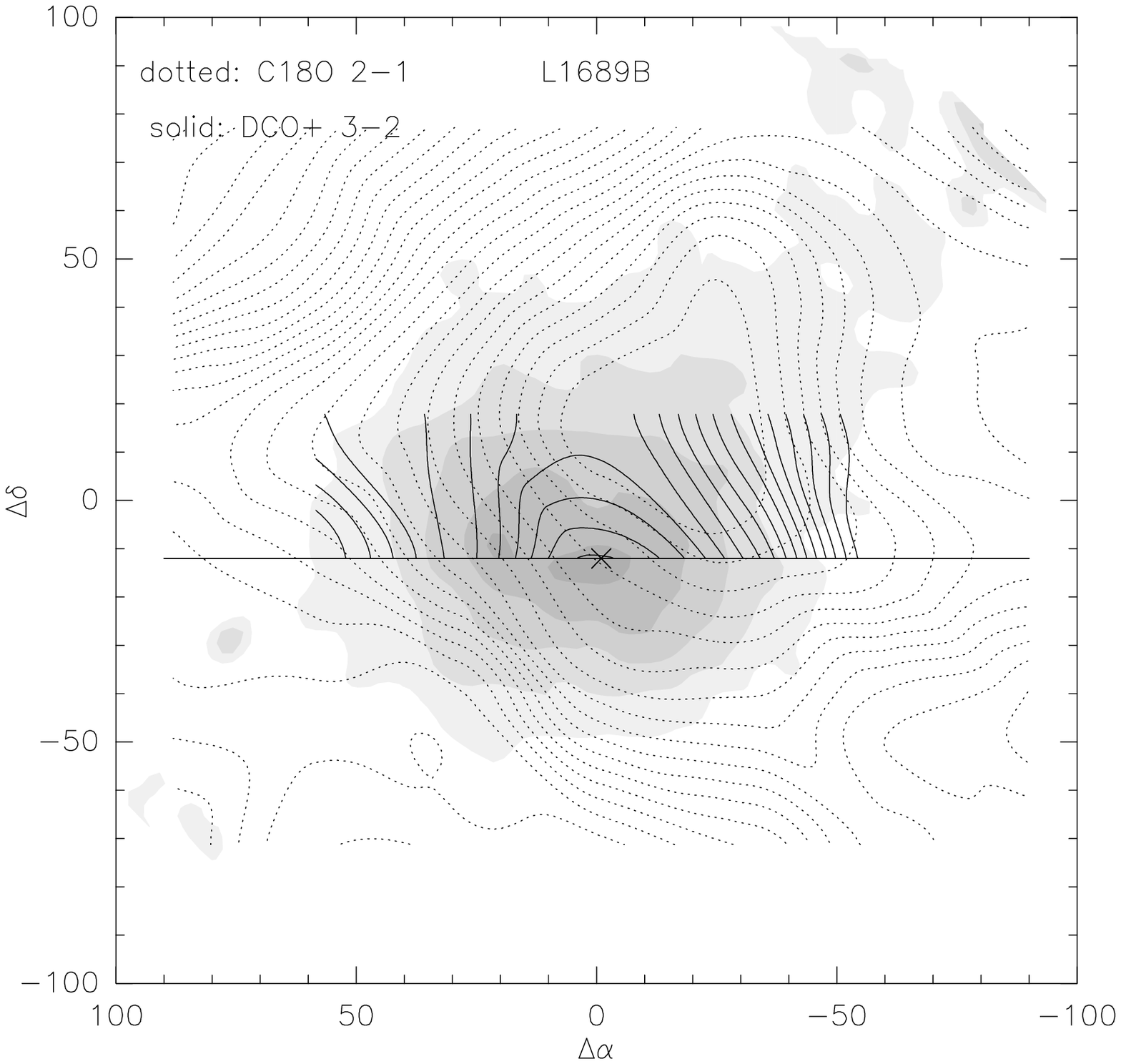}
\figcaption{
}
\end{figure}

\begin{figure}
\figurenum{3}
\plotone{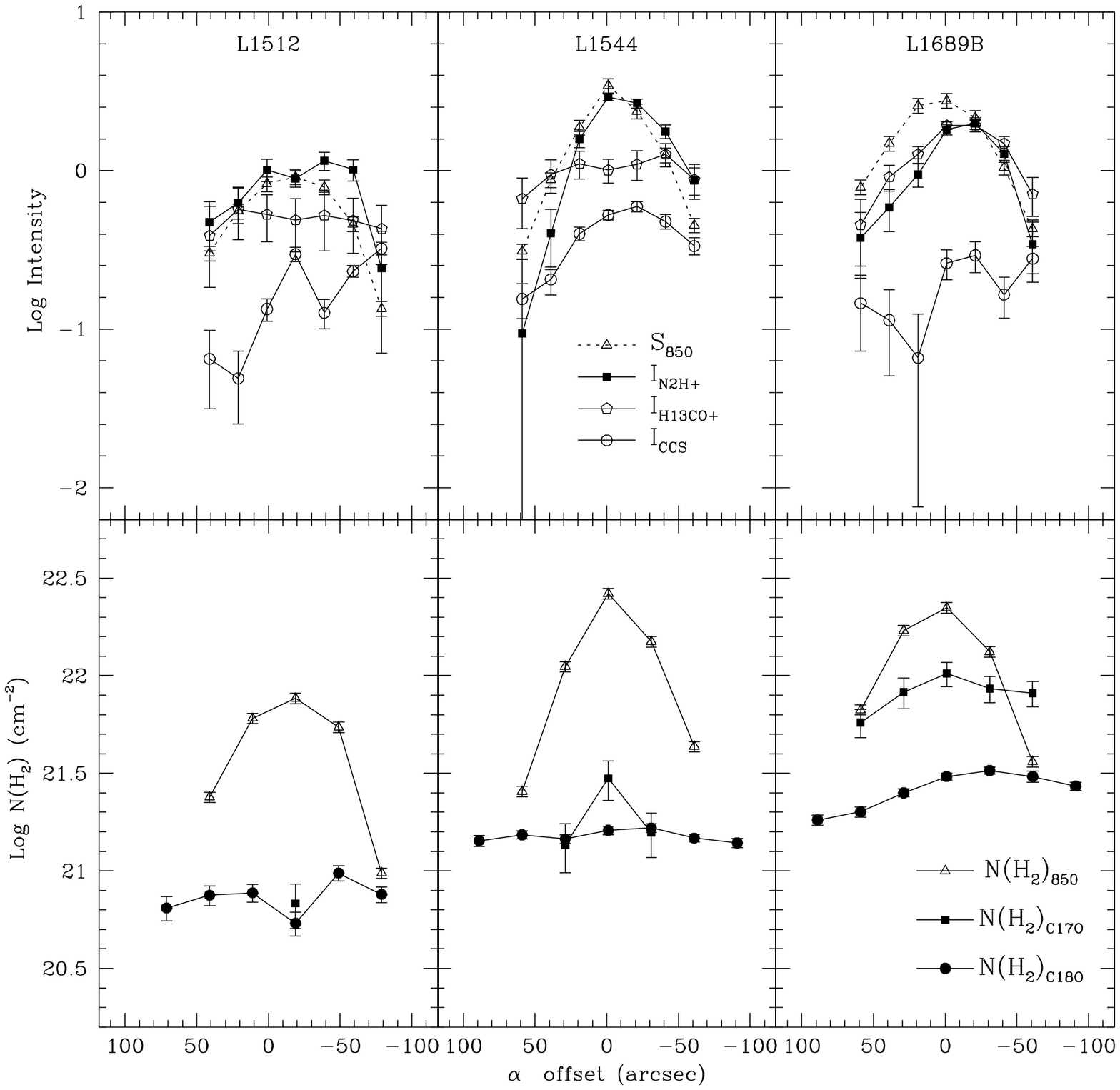}
\figcaption{
These plots in upper panels compare the integrated intensities of CCS
$\rm N_J=4_3-3_2$, $\rm H^{13}CO^+$ $\rm J=1-0$, and $\rm N_2H^+$ $\rm J=1-0$ 
lines with $S_{850}$ through the cuts marked in Figure 2.
Here, $S_{850}$ is shifted by 0.9, and $I_{\rm H^{13}CO^+}$ is shifted by 
0.4 in the logarithmic scale.
These plots in lower panels compare the column densities of hydrogen molecules
calculated from the 850 $\rm \mu m$ dust continuum emission,
$\rm C^{18}O$ $\rm J=2-1$ line (without the correction of $\tau$), and
$\rm C^{17}O$ $\rm J=2-1$ line (with correction of $\tau$)
through the cuts marked in Figure 2.
The errors in molecular intensities only account for the rms noises of spectra 
so that the error bars show the minimum errors.
In $\rm N(H_2)$ derived from \c17o of L1544 and L1689B, the error in fitting
the hyperfine structure is also included.
}
\end{figure}

\begin{figure}
\figurenum{4}
\plotone{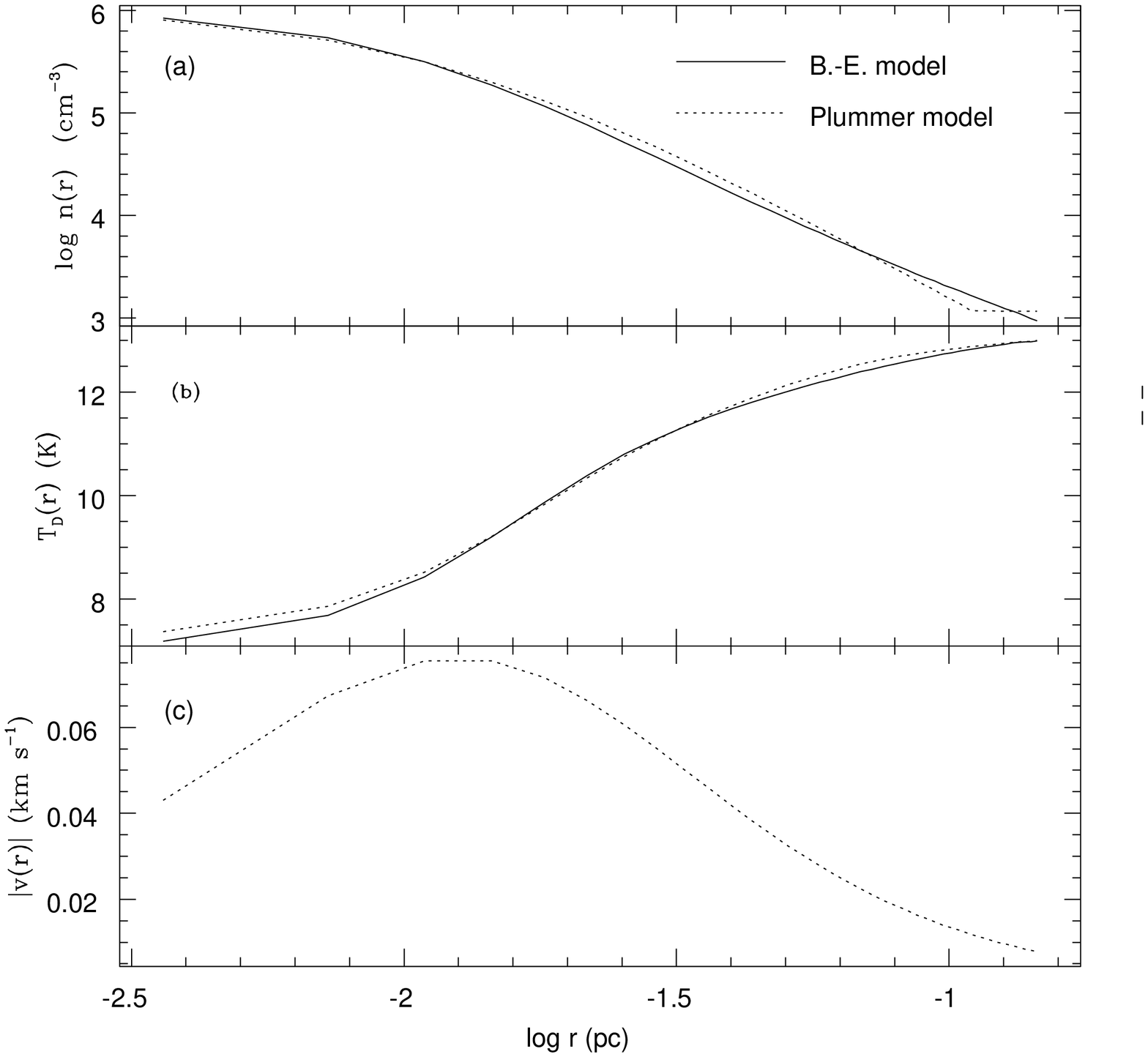}
\figcaption{
The comparison between physical parameters of the Bonnor-Ebert model and the 
Plummer-like model: (a) the density distribution, (b) the temperature 
distribution, and (c) the velocity distribution. In the density structure of
Plummer-like model, we forced the lower limit of density to be $10^3$ cm$^{-3}$.
}

\end{figure}
\begin{figure}
\figurenum{5}
\plotone{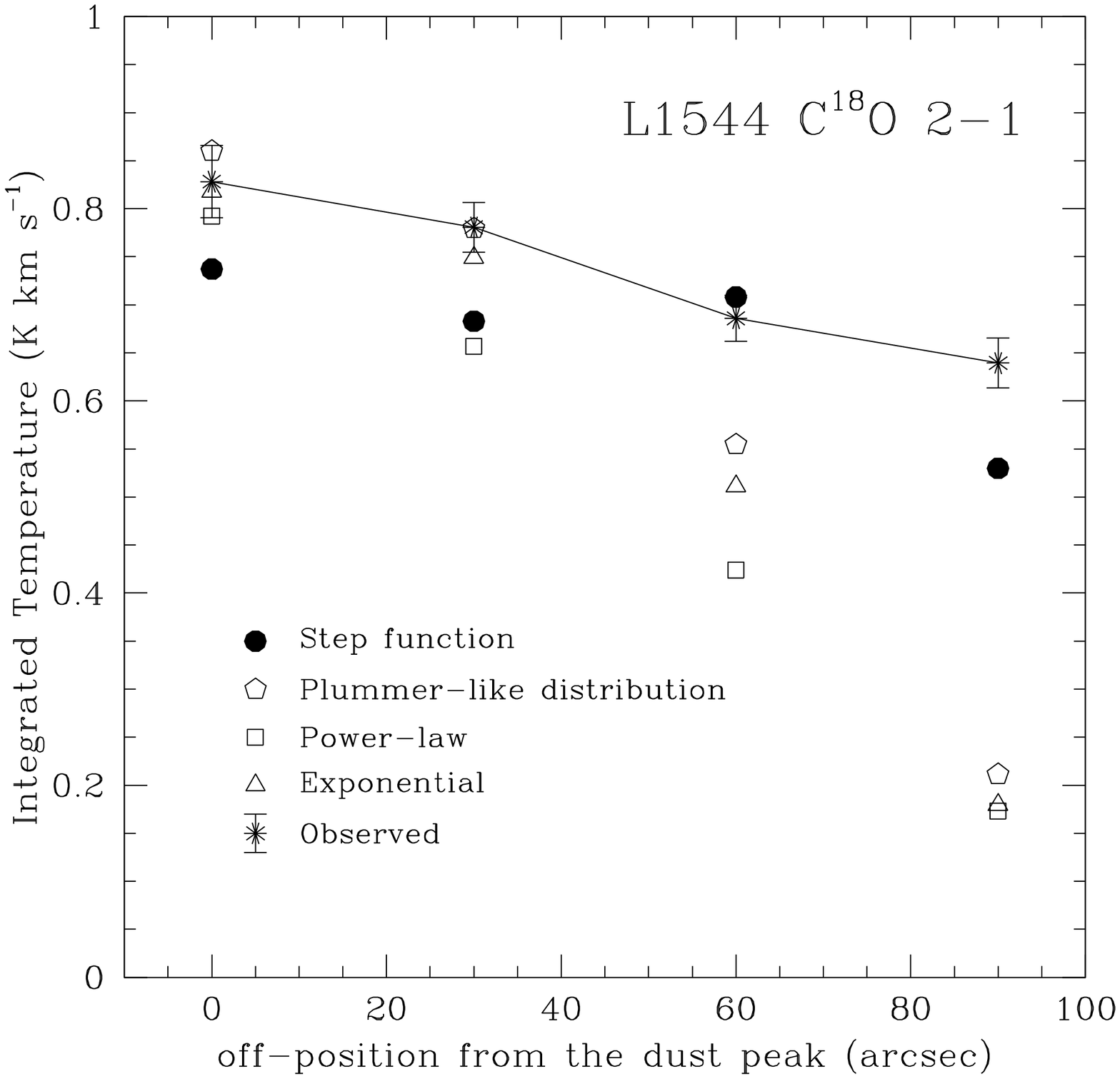}
\figcaption{
The comparison of different functional forms of depletion (Table 6).
The integrated interval of velocity is from 6.5 to 8 km s$^{-1}$. 
The step function fits best the distribution of the observed integrated 
temperatures.  The line profiles of modeling the step function is shown in 
Figure 7.a. 
The integrated temperatures of the model of the step function are smaller than 
the observed ones because the model does not fit the high velocity wing parts 
that increase the integrated temperatures in the observed lines.
}
\end{figure}

\begin{figure}
\figurenum{6.a}
\plotone{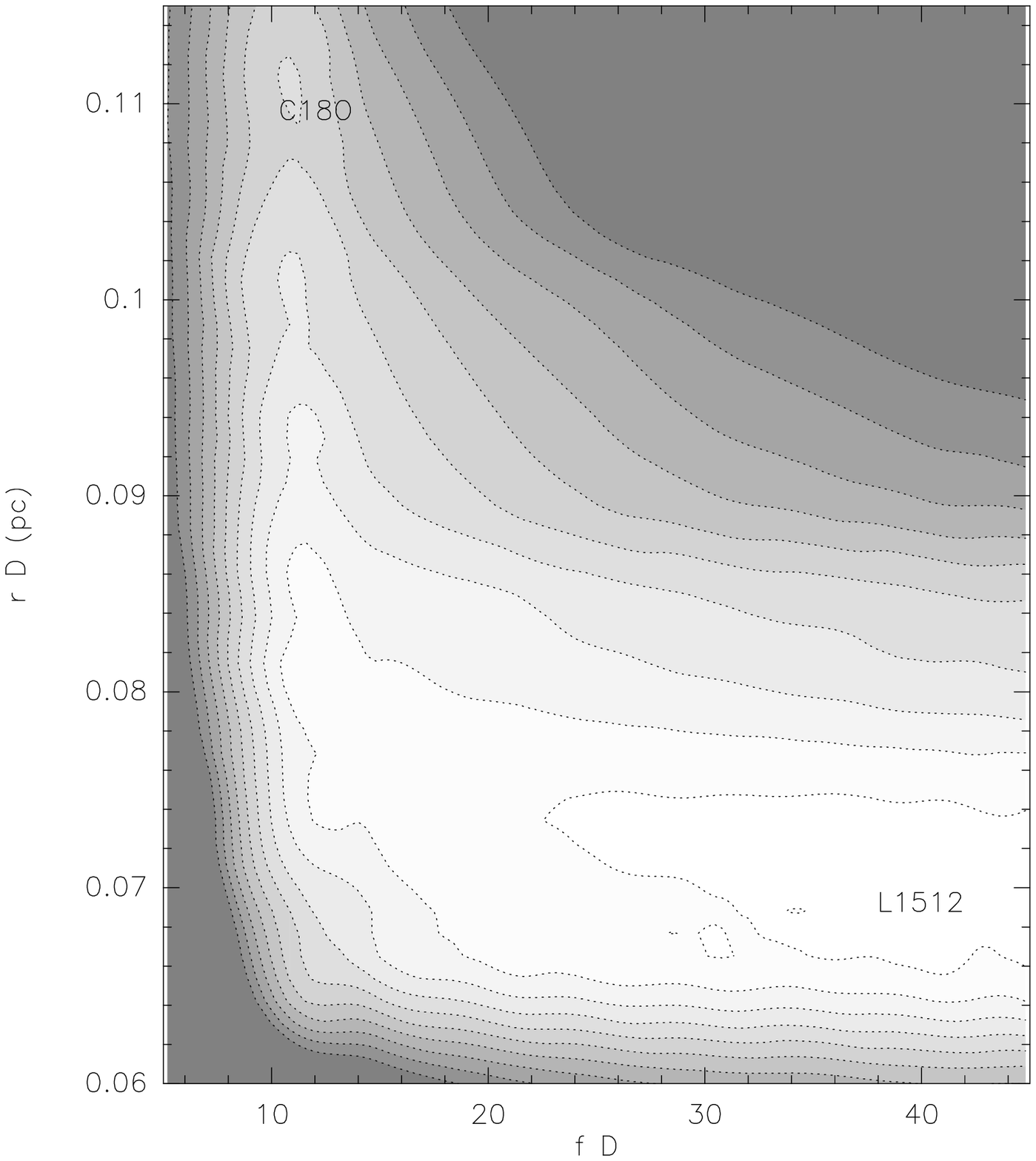}
\figcaption{
The distribution of the reduced $\chi ^2$ of models that have different
r$\rm _D$ and f$\rm _D$ in C$^{18}$O 2$-$1 and 3$-$2 toward L1512. 
The contour levels are from 5 to 95 increasing by the interval of 10. 
The $\chi ^2$ has been calculated by weighting the difference between the 
modeled and observed integrated temperatures with the errors in the 
integrated temperatures of the observed lines. 
}
\end{figure}

\begin{figure}
\figurenum{6.b}
\plotone{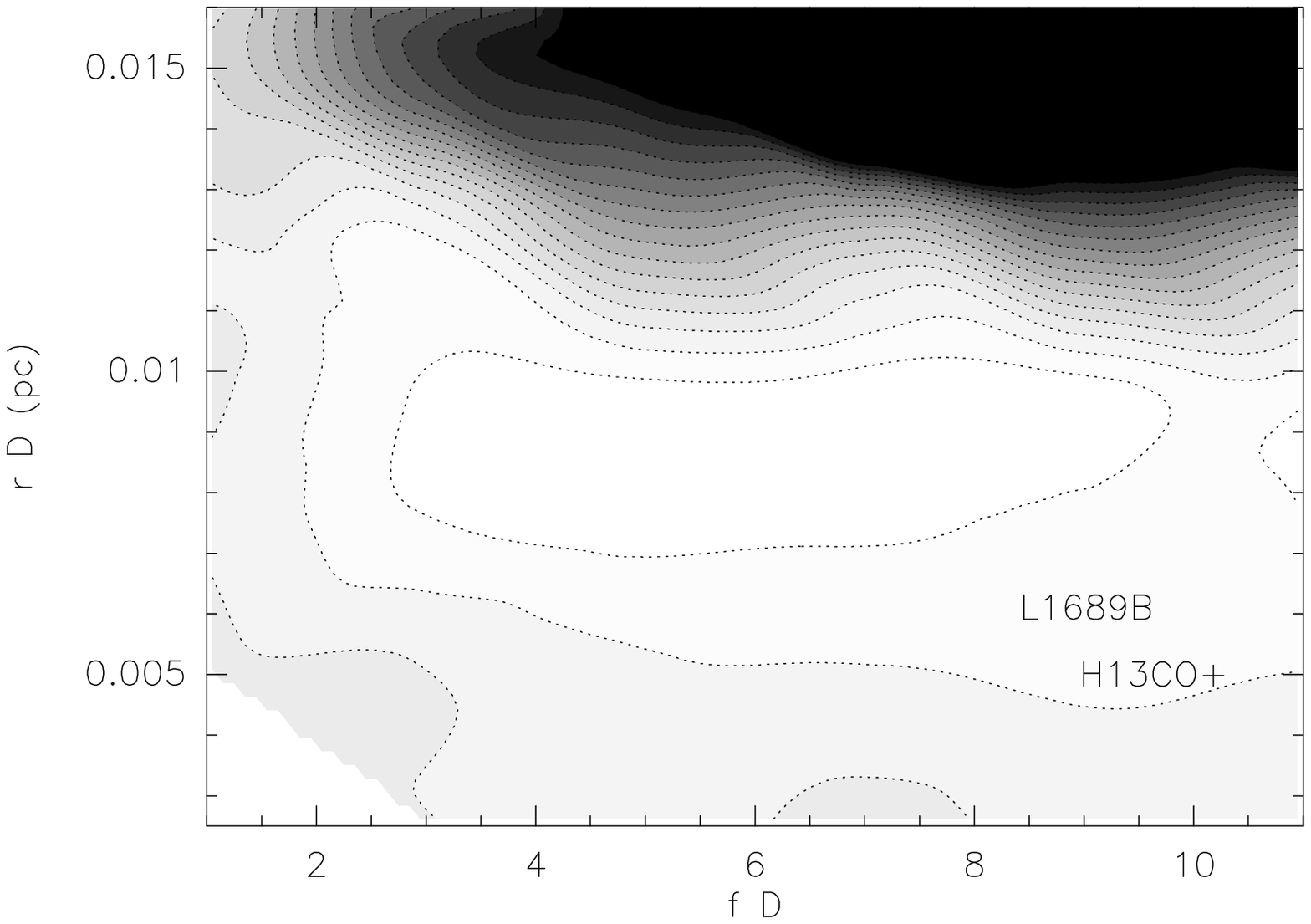}
\figcaption{
The distribution of the reduced $\chi ^2$ of models that have different
r$\rm _D$ and f$\rm _D$ in H$^{13}$CO$^+$ 1$-$0 and 3$-$2 toward L1689B.
The contour levels are from 3 to 33 increasing by the interval of 2. 
}
\end{figure}

\begin{figure}
\figurenum{7.a}
\plotone{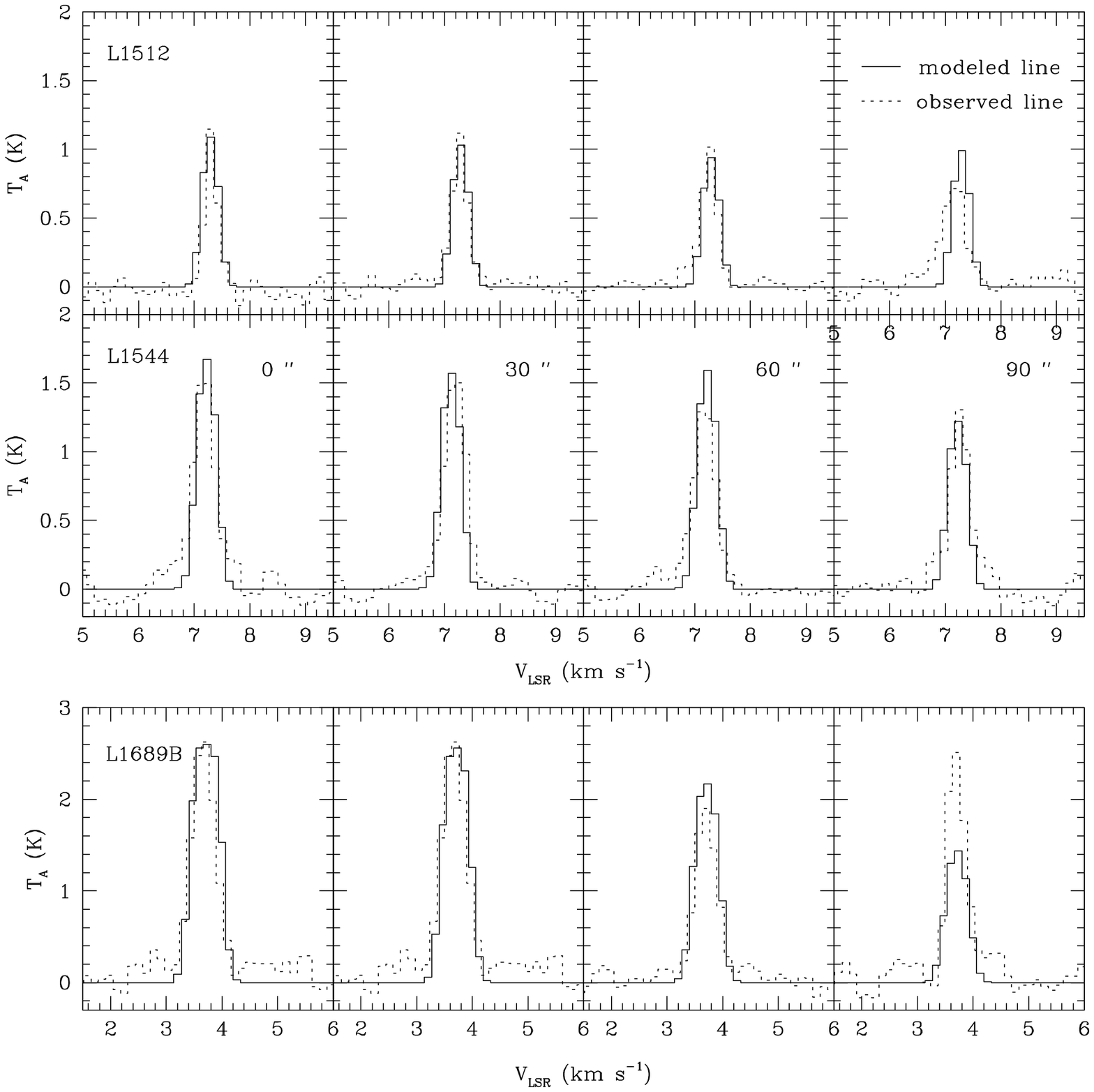}
\figcaption{
The results of MC modeling in \dc18o $\rm J=2-1$. \dc18o $\rm J=2-1$ and 
$\rm J=3-2$ lines have been modeled simultaneously, and we chose the abundance 
structure that fitted both of the lines.
In each panel, the solid line and the dashed line indicate the modeled 
and the observed line profiles, respectively. The arcseconds marked in the 
middle panels represent the angular distance from the dust peak of each core.
}
\end{figure}

\begin{figure}
\figurenum{7.b}
\plotone{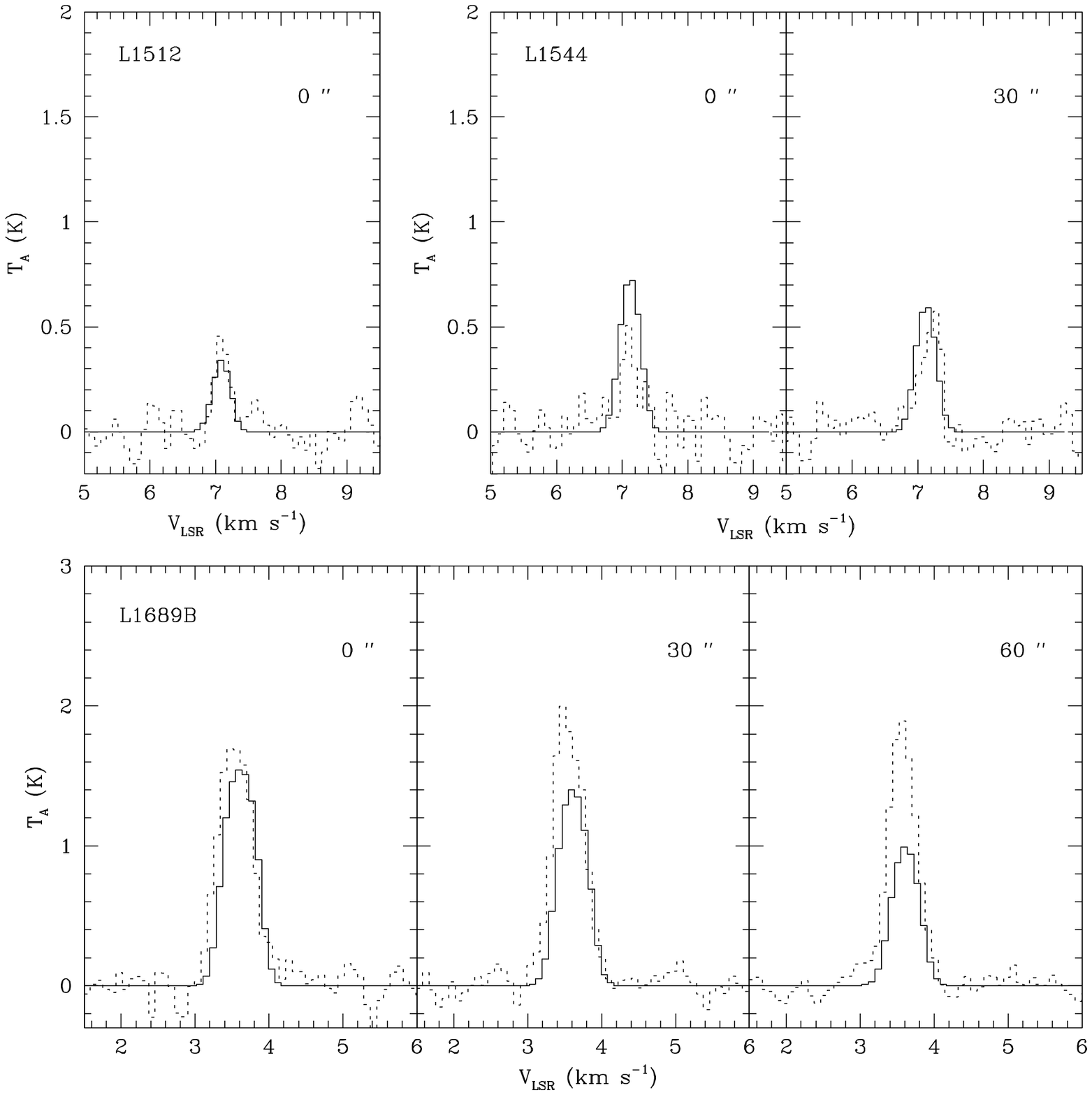}
\figcaption{
The results of MC modeling in \dc18o $\rm J=3-2$.
}
\end{figure}

\begin{figure}
\figurenum{7.c}
\plotone{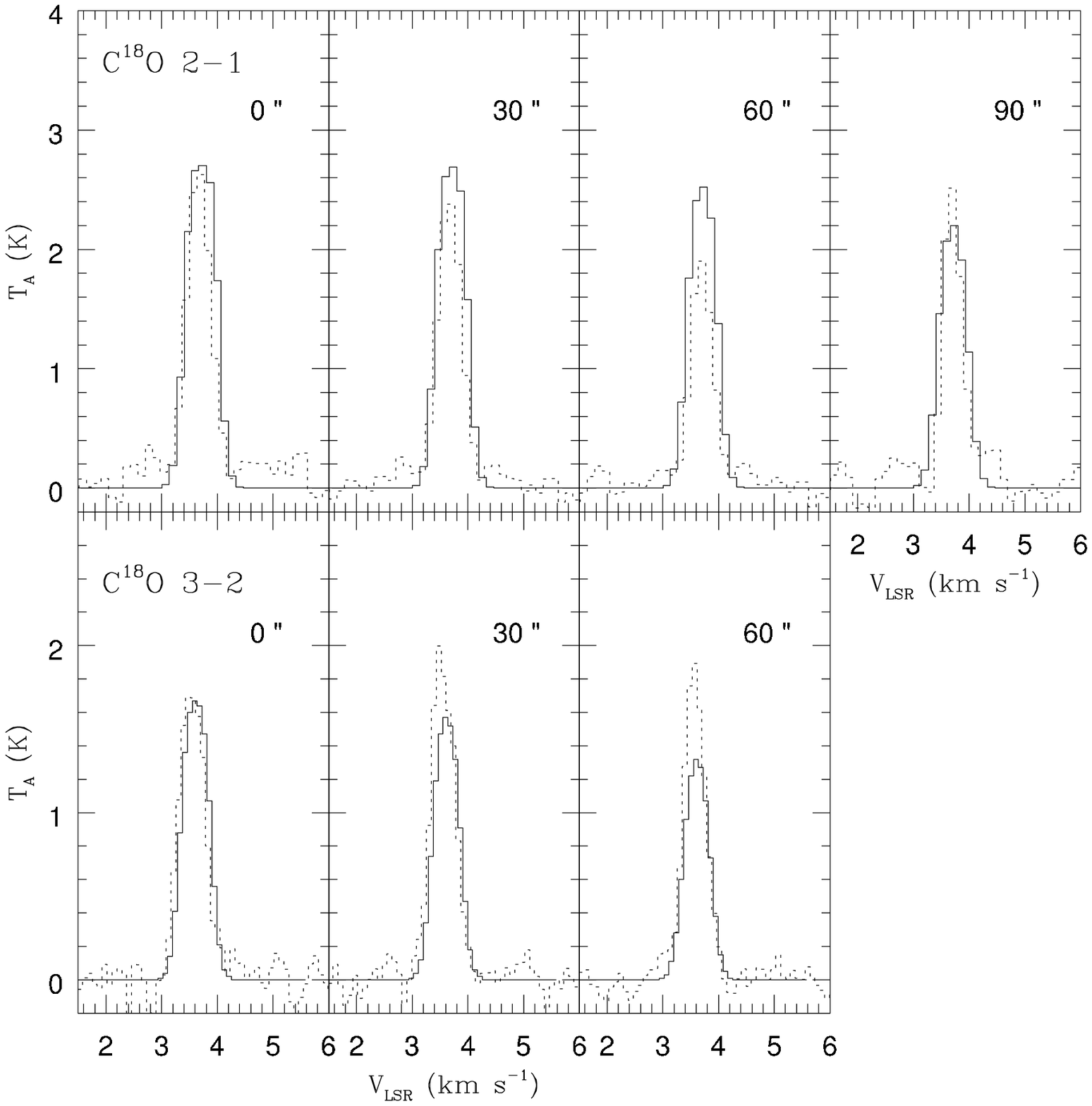}
\figcaption{
The results of MC modeling of L1689B in C$^{18}$O lines including 
a warm envelope.
}
\end{figure}

\begin{figure}
\figurenum{8.a}
\plotone{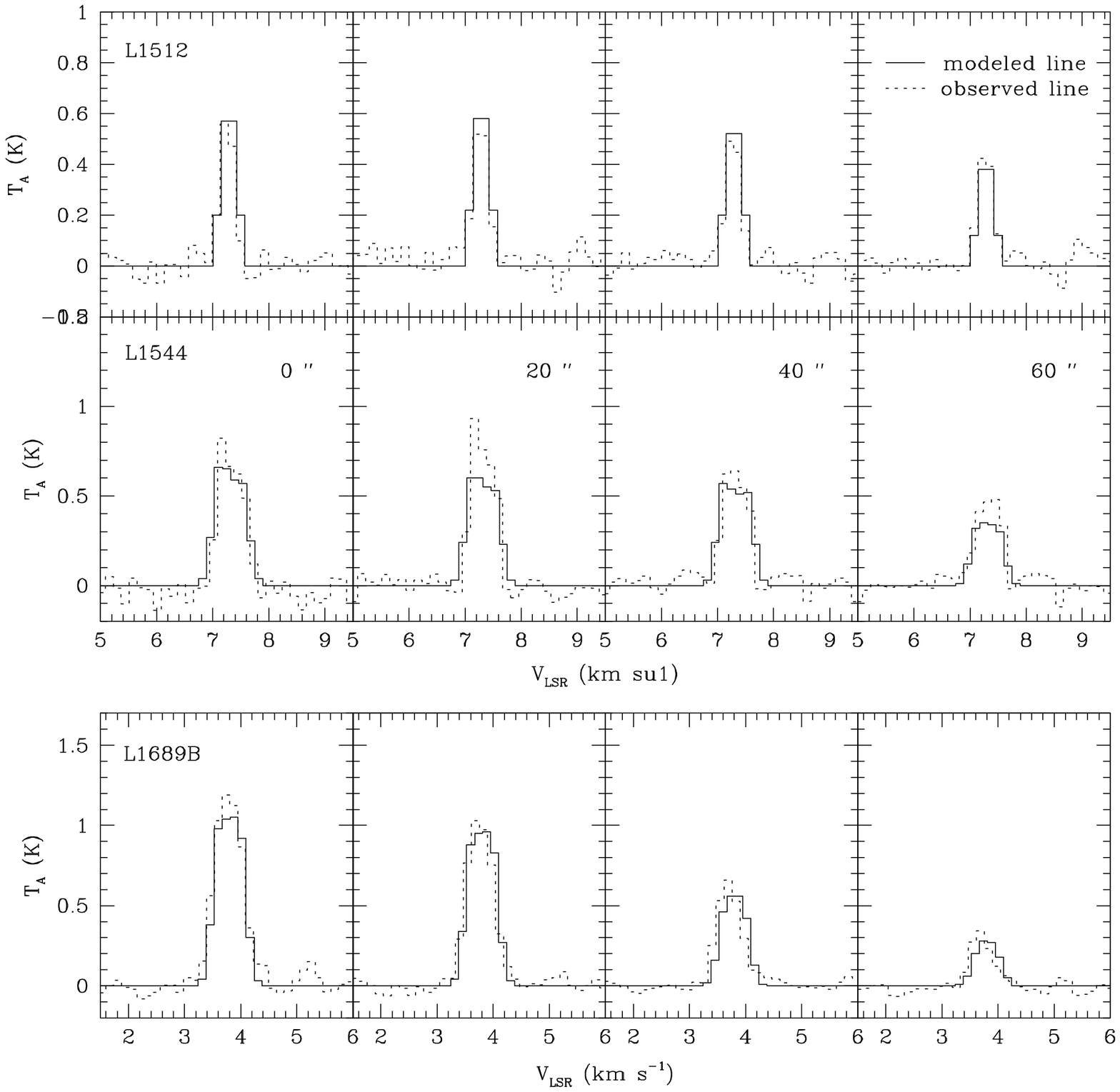}
\figcaption{
The results of MC modeling in $\rm H^{13}CO^+$ $\rm J=1-0$.  $\rm H^{13}CO^+$ 
$\rm J=1-0$ and $\rm J=3-2$ lines have been modeled simultaneously, and we 
chose the abundance structure that fitted both of the lines.
In order to model these line profiles, in L1512 and L1689B, we used the 
density and temperature profiles given by Bonnor-Ebert dust models 
(Evans et al. 2001), but we calculated the structures of density
and temperature in L1544 combining the empirical dynamics model of 
Whitworth and Ward-Thompson (2001) and dust radiative transfer code.
}
\end{figure}

\begin{figure}
\figurenum{8.b}
\plotone{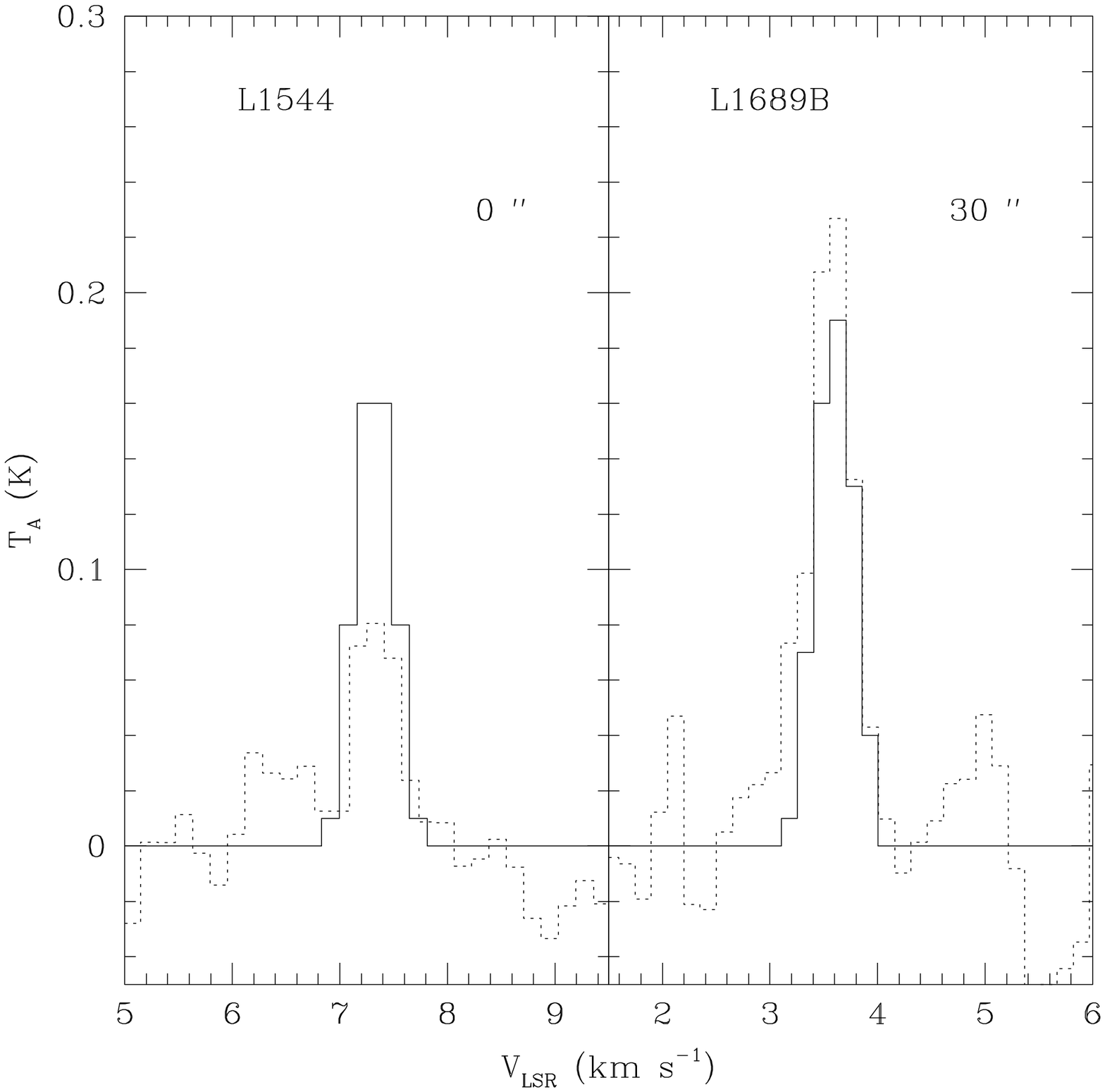}
\figcaption{
The results of MC modeling in $\rm H^{13}CO^+$ $\rm J=3-2$. 
}
\end{figure}

\begin{figure}
\figurenum{9}
\plotone{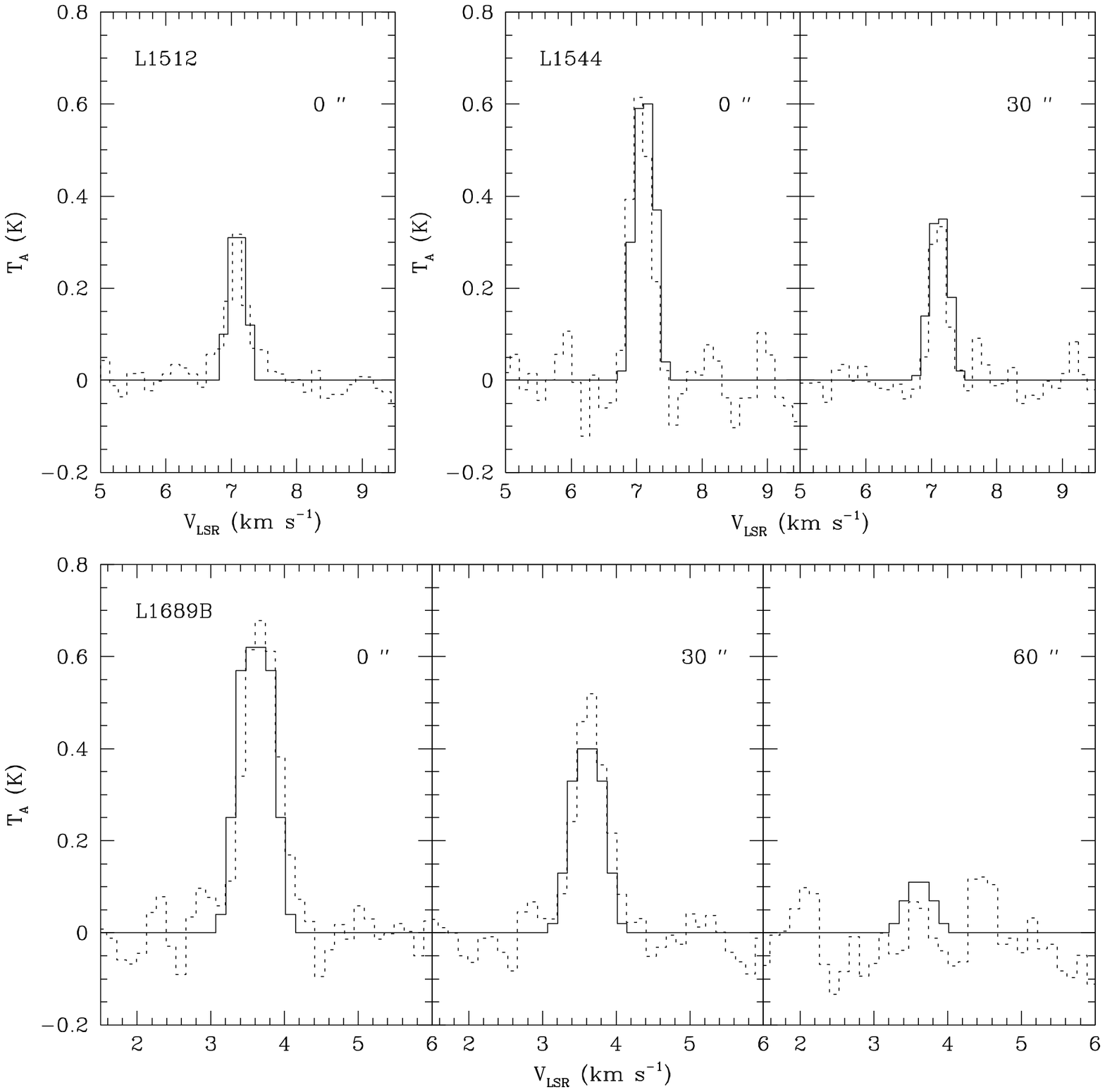}
\figcaption{ 
The results of MC modeling in $\rm DCO^+$ $\rm J=3-2$ line.
}
\end{figure}

\begin{figure}
\figurenum{10}
\plotone{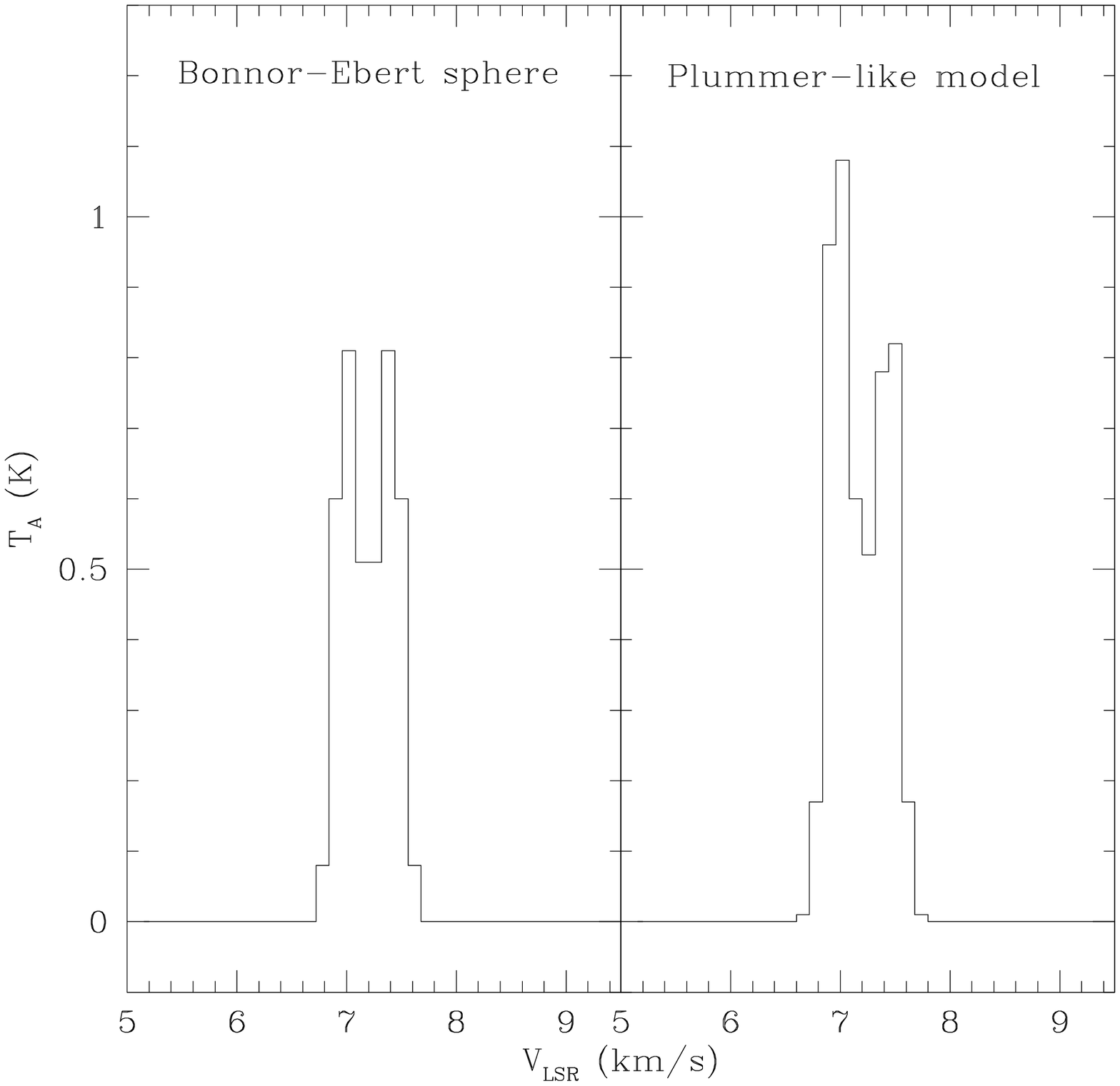}
\figcaption{ 
The comparison of MC modeling in $\rm HCO^+$ $\rm J=3-2$ line of L1544 for 
two different physical models: Bonnor-Ebert sphere and Plummer-like model
(see Figure 4).
Both have the same abundance structure, but only the Plummer-like model, with 
an infall velocity structure, produces the blue asymmetry shown in 
observed line profiles.
}
\end{figure}

\begin{figure}
\figurenum{11}
\plotone{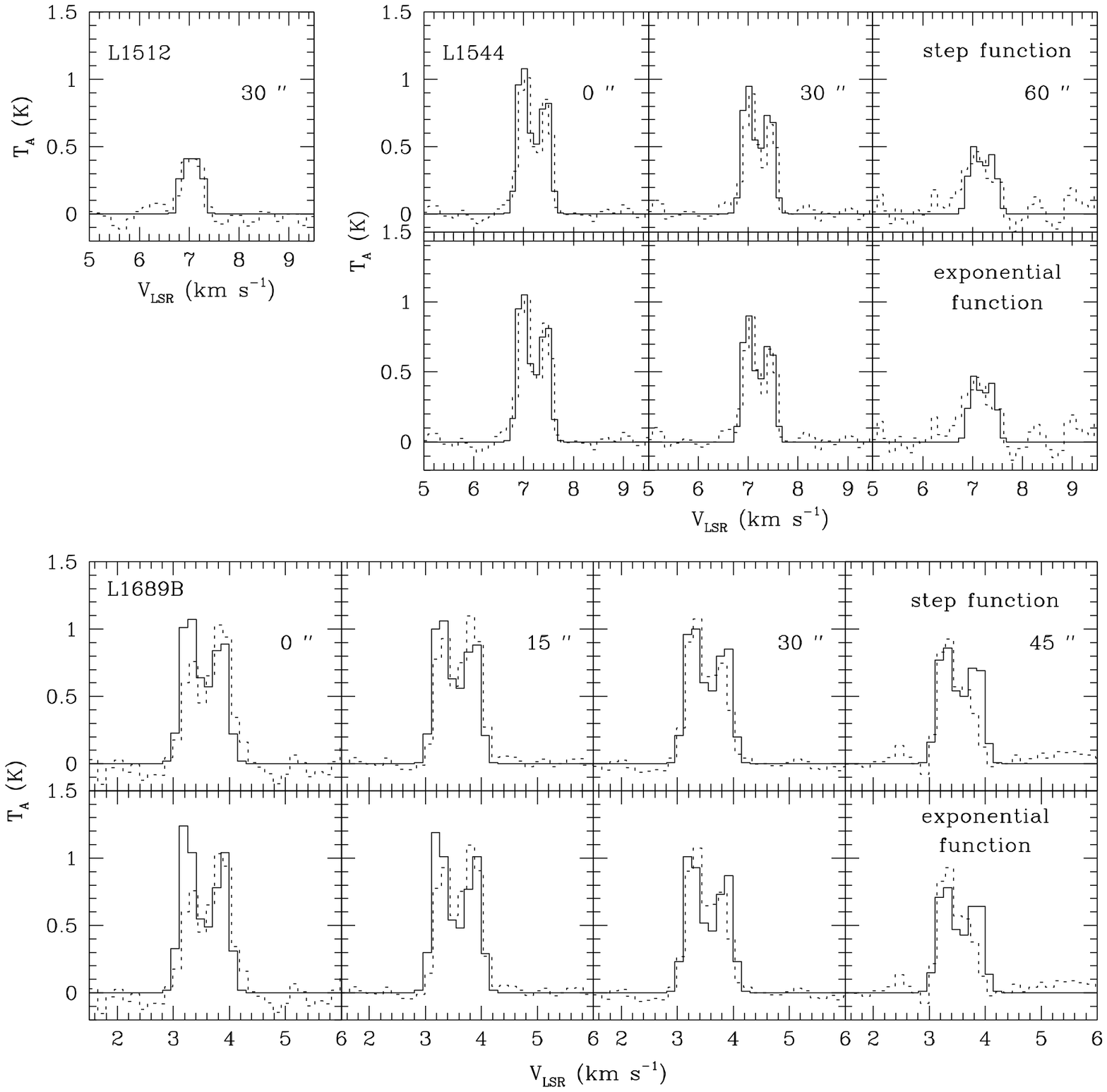}
\figcaption{ 
The results of MC modeling in $\rm HCO^+$ $\rm J=3-2$ line.
In L1544 and L1689B, the upper and the lower panels show the results by 
the step function and the exponential function of $\rm HCO^+$ abundance,
respectively. In L1512, we modeled the line profile using a constant abundance. 
}
\end{figure}

\begin{figure}
\figurenum{12}
\plotone{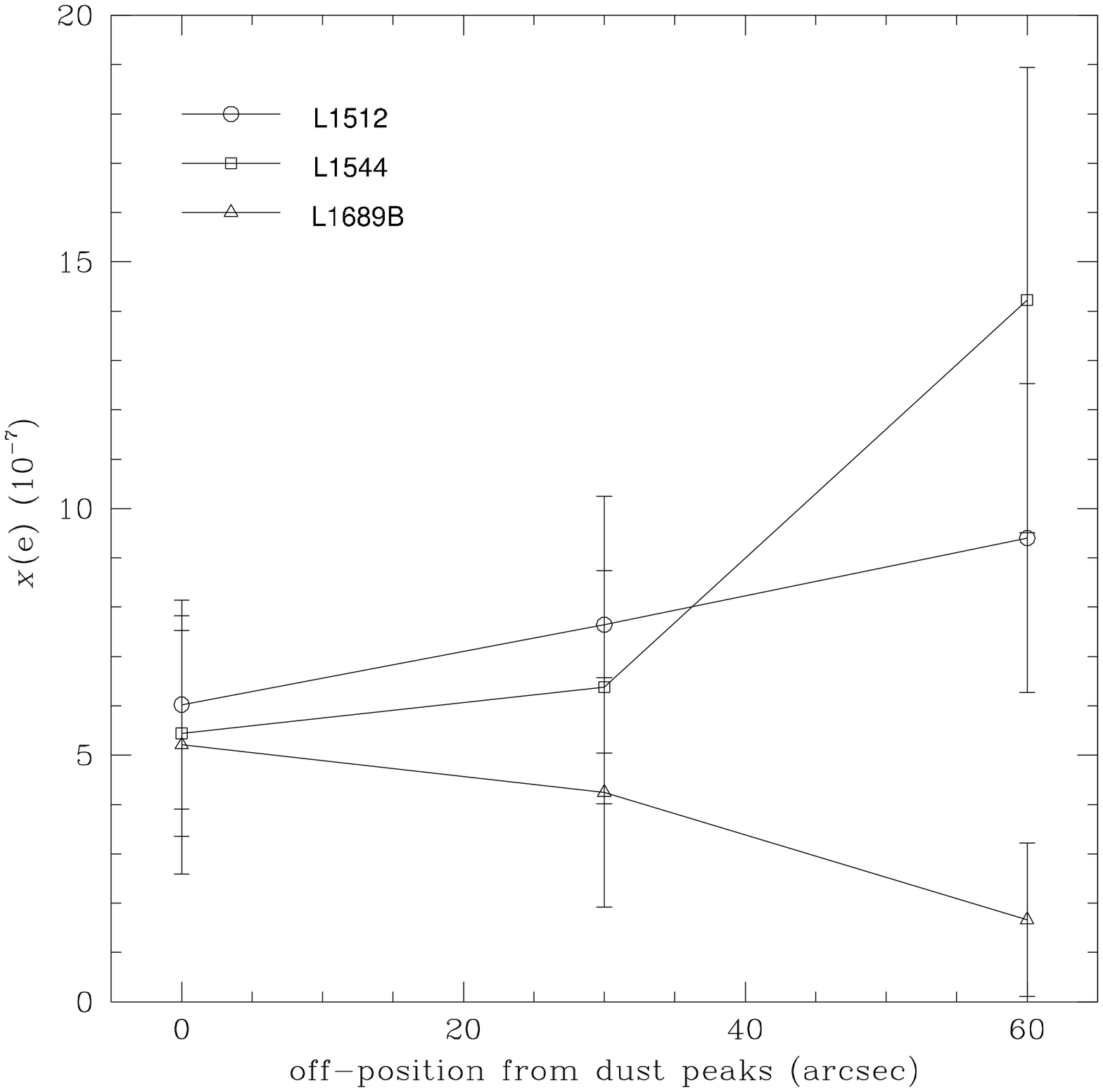}
\figcaption{
The ionization fractions in three PPCs. The error bars represent the 
amount of uncertainty in the ionization fraction 
when the ratio of $\rm DCO^+/HCO^+$ and 
the depletion fraction of CO molecule changes up to 30 \% of the calculated
values in each point.
}
\end{figure}

\clearpage

%%%%%%%%%%%%%%%%%%%table1%%%%%%%%%%%%%%%%%%%%%%%%%%%%%%%%%%%%%%%%
\begin{deluxetable}{lccccc}
\tablecolumns{5}
\footnotesize
\tablecaption{\bf List of sources}
\tablewidth{0pt} 
\tablehead{
\colhead{Name}                  &
\colhead{R.A. (1950.0)}  &
\colhead{Dec. (1950.0)}  &
\colhead{$V_{LSR}$}    & 
\colhead{distance}             & 
\colhead{offset of dust peak}  \\            
\colhead{} &  
\colhead{ hh mm ss }      &
\colhead{\degree~~ \am~~ \as}    &  
\colhead{$\rm km~s^{-1}$} &              
\colhead{pc}     &          
\colhead{arcsec} 
}
\startdata 
 L1512    & 05 00 54.4  &  32 39 37   &  7.09  & 140& ($-19$, $-26$)\\
 L1544    & 05 01 13.1  &  25 06 36   &  7.12  & 140& ($-1$, 4)\\
 L1689B   & 16 31 47.0  & $-24$ 57 22   &  3.61  & 125& ($-1$, $-12$)\\
\enddata
\end{deluxetable}

%%%%%%%%%%%%%%%%%%%table2%%%%%%%%%%%%%%%%%%%%%%%%%%%%%%%%%%%%%%%%
\begin{deluxetable}{lccccc}
\tablecolumns{6}
\footnotesize
\tablecaption{\bf List of observed lines}
\tablewidth{0pt} 
\tablehead{
\colhead{Molecule}  &
\colhead{Transition}  &
\colhead{Frequency}  &
\colhead{Beam Width}  &
\colhead{Velocity Resolution} &
\colhead{$\eta_{\rm mb}$\tablenotemark{d}}\\ 
\colhead{}                        &
\colhead{}    &  
\colhead{MHz} &  
\colhead{arcsec} &  
\colhead{$\rm km~s^{-1}$} &               
\colhead{L1512/L1544/L1689B}
}
\startdata 
 \c17o           & $\rm J=2-1$\tablenotemark{a}        &  224714.385 &  33  & 0.17 & 0.65/0.65/0.80\\
 \dc18o          & $\rm J=2-1$\tablenotemark{a}        &  219560.357 &  34  & 0.15 & 0.69/0.66/0.60\\
                 & $\rm J=3-2$\tablenotemark{a}        &  329330.507 &  26  & 0.10 & 0.59/0.59/0.63\\
$\rm HCO^+$      & $\rm J=3-2$\tablenotemark{a,c}      &  267557.620 &  26  & 0.15 & 0.49/0.66/0.65 \\
$\rm H^{13}CO^+$ & $\rm J=1-0$\tablenotemark{b}        &   86754.330 &  18  & 0.14 & 0.49/0.49/0.49\\
                 & $\rm J=3-2$\tablenotemark{a,c}      &  260255.617 &  26  & 0.16 & none/0.49/0.65 \\
 \dcop           & $\rm J=3-2$\tablenotemark{a}        &  216112.605 &  35  & 0.15 & 0.65/0.69/0.60\\
 CCS             & $\rm N_J=4_3-3_2$\tablenotemark{b}  &   45379.033 &  37  & 0.27 & 0.77/0.77/0.77\\
%                 & $\rm N_J=8_7-7_6$\tablenotemark{b}  &   93870.107 &  17  &      & 0.50/0.50/0.50\\
$\rm N_2H^+$     & $\rm J=1-0$\tablenotemark{b}        &   93173.809 &  17  & 0.13 & 0.50/0.50/0.50\\
%                 & $\rm J=3-2$\tablenotemark{a}        &  279511.000 &  27  &      & 0.53\\
\enddata
\tablenotetext{a}{observed with CSO 10.4 m telescope}
\tablenotetext{b}{observed with Nobeyama 45 m telescope}
\tablenotetext{c}{Gregersen and Evans, 2000}
\tablenotetext{d}{$\eta_{mb}\equiv T_A^*/T_R^*$, where $T_R^*$ is the radiation
temperature, and $T_R=T_R^*$ only if a source fills the beam (Kutner and Ulich
1981).}
\end{deluxetable}

%%%%%%%%%%%%%%%%%%%table3%%%%%%%%%%%%%%%%%%%%%%%%%%%%%%%%%%%%%%%%
%\begin{deluxetable}{lccc}
%\tablecolumns{4}
%\footnotesize
%\tablecaption{\bf The main beam efficiency and the Pointing corrections (230 GHz Receiver)}
%\tablewidth{0pt} 
%\tablehead{
%\colhead{Date}                  &
%\colhead{Main Beam efficiency}  &
%\colhead{FAZO}       &
%\colhead{FZAO}                   \\
%\colhead{}                        &
%\colhead{}    &  
%\colhead{arcsecond} &  
%\colhead{arcsecond}                  
%}
%\startdata 
% Jun 1998 & 0.70  & -50.4 ($\pm$ 6.8) &  33.6 ($\pm$ 2.7) \\
% Dec 1998 & 0.66  & -70.2 ($\pm$ 2.4) & -13.9 ($\pm$ 3.4) \\
% Jul 1999 & 0.62  & -81.2 ($\pm$ 5.6) & -39.0 ($\pm$ 5.4) \\
% Dec 1999 & 0.69  & -64.4 ($\pm$ 2.5) & -12.3 ($\pm$ 4.4) \\
% Jun 2000 & 0.81  & -64.4 ($\pm$ 2.5) & -12.3 ($\pm$ 4.4) \\
% Dec 2000 & 0.65  & -64.4 ($\pm$ 2.5) & -12.3 ($\pm$ 4.4) \\
%\enddata
%\end{deluxetable}

%%%%%%%%%%%%%%%%%%%table3%%%%%%%%%%%%%%%%%%%%%%%%%%%%%%%%%%%%%%%%
\begin{deluxetable}{lccc}
\tablecolumns{4}
\footnotesize
\tablecaption{\bf Parameters for molecules}
\tablewidth{0pt} 
\tablehead{
\colhead{Isotope}                  &
\colhead{Rotational constant}  &
\colhead{Dipole moment}      &             
\colhead{Abundance}                   \\
\colhead{}                        &
\colhead{$B$ (MHz)}    &  
%\colhead{B (MHz)\tablenotemark{a}}    &  
%\colhead{\mu}  & 
\colhead{$\mu$ (Debye)}  & 
\colhead{$X$($=n_i/n_{H_2}$)}   
}
\startdata 
 \dc18o          & 54891.423  &  0.11079 & $4.8\times 10^{-7}$   \\
 \c17o           & 56179.982  &  0.11034 & $1.5\times 10^{-7}$   \\
%$\rm H^{13}CO^+$ & 43377.3  &  3.93 & $10^{-10}$   \\
%$\rm DCO^+$      & 36019.8  &  3.93 & $3.5\times 10^{-10}$   \\
\enddata
\end{deluxetable}

%%%%%%%%%%%%%%%%%%%table4%%%%%%%%%%%%%%%%%%%%%%%%%%%%%%%%%%%%%%%%
\begin{deluxetable}{lc}
\tablecolumns{2}
\footnotesize
\tablecaption{Optical depth of $\rm C^{17}O$ $\rm J=2-1$ line at center position
}
\tablewidth{0pt}
\tablehead{
\colhead{Source}                  &
\colhead{Total optical depth}
}
\startdata
 L1512   &  0.10 ($\pm$2.17)   \\
 L1544   &  1.03 ($\pm$0.42)   \\
 L1689B  &  0.79 ($\pm$0.24)   \\
\enddata
\end{deluxetable}

%%%%%%%%%%%%%%%%%%%table5%%%%%%%%%%%%%%%%%%%%%%%%%%%%%%%%%%%%%%%%
\begin{deluxetable}{lc}
\tablecolumns{2}
\footnotesize
\tablecaption{Depletion factor of CO molecule at center position}
\tablewidth{0pt}
\tablehead{
\colhead{Source}                  &
\colhead{$N$$\rm (H_2)_{S_{850}}$/$N$$\rm (H_2)_{C^{17}O\tablenotemark{a}}$}
}
\startdata
 L1512   &  11  \\
 L1544   &  9 \\
 L1689B  &  2 \\
\enddata
\tablenotetext{a}{We have corrected for the optical depth effect to calculate the
column density.}
\end{deluxetable}

%%%%%%%%%%%%%%%%%%%table6%%%%%%%%%%%%%%%%%%%%%%%%%%%%%%%%%%%%%%%%
\begin{deluxetable}{lcc}
\tablecolumns{3}
\footnotesize
\tablecaption{The $\chi ^2$ in modeling the C$^{18}$O 2$-$1 lines of L1544 by
the different functional forms of depletion}
\tablewidth{0pt}
\tablehead{
\colhead{Function}                  &
\colhead{$X$}&
\colhead{$\chi ^2$} 
}
\startdata
 Step   &  $X_0$\tablenotemark{a}~~ (r $>$ 0.045 pc) & 10\\
        &  $X_0/25$ (r $<$ 0.045 pc) & \\
 Plummer-like & $X_0\left[\frac{r}{\sqrt{r^2+(0.05)^2}}\right]^2$ & 76 \\
 Power-law & $X_0 (\frac{r}{0.15})^1$ & 117 \\
 Exponential & $X_0 exp(-\frac{0.04}{r})$ & 93 \\ 
\enddata
\tablenotetext{a}{$X_0=4.82\times10^{-7}$} 
\end{deluxetable}

%%%%%%%%%%%%%%%%%%%table7%%%%%%%%%%%%%%%%%%%%%%%%%%%%%%%%%%%%%%%%
\begin{deluxetable}{lcccc}
\tablecolumns{5}
\footnotesize
\tablecaption{The parameters of best-fit models in $\rm C^{18}O$ lines}
\tablewidth{0pt}
\tablehead{
\colhead{Source}                  &
\colhead{$X_0$\tablenotemark{a}}&
\colhead{$r_D$\tablenotemark{a}}&
\colhead{$f_D$\tablenotemark{a}}&
\colhead{$V_{\rm microturbulence}$}\\ 
\colhead{}&
\colhead{}&
\colhead{pc}&
\colhead{}&
\colhead{$\rm km~s^{-1}$}
}
\startdata
 L1512   &  $4.82\times10^{-7}$ & 0.075 & 25 & 0.15 \\
 L1544   &  $4.82\times10^{-7}$ & 0.045  & 25& 0.17 \\
 L1689B\tablenotemark{b}  &  $4.82\times10^{-7}$ & 0.03  &  3 & 0.20\\
\enddata
\tablenotetext{a}{$X=X_0~~(r\ge r_D)$ and $X=X_0/f_D~~(r<r_D)$ where $X_0$
is undepleted abundance, $r_D$ is the radius within which a molecule is 
depleted, and $f_D$ is the fraction of depletion.}
\tablenotetext{b}{We could not find a good model to fit all line profiles in
all observed positions in L1689B.}  
\end{deluxetable}

%%%%%%%%%%%%%%%%%%%table8%%%%%%%%%%%%%%%%%%%%%%%%%%%%%%%%%%%%%%%%
\begin{deluxetable}{lcccc}
\tablecolumns{5}
\footnotesize
\tablecaption{The parameters of best-fit models in $\rm H^{13}CO^+$ lines}
\tablewidth{0pt}
\tablehead{
\colhead{Source}                  &
\colhead{$X_0$}&
\colhead{$r_D$}&
\colhead{$f_D$}&
\colhead{$V_{\rm microturbulence}$} \\
\colhead{}&
\colhead{}&
\colhead{pc}&
\colhead{}&
\colhead{$\rm km~s^{-1}$}
}
\startdata
 L1512   &  $3.0\times10^{-10}$ & 0.021 & 25 & 0.07\\
 L1544   &  $5.0\times10^{-10}$ & 0.026 & 20 & 0.20\\
 L1689B  &  $2.2\times10^{-10}$ & 0.012 &  5 & 0.20\\
\enddata
\end{deluxetable}

%%%%%%%%%%%%%%%%%%%table9%%%%%%%%%%%%%%%%%%%%%%%%%%%%%%%%%%%%%%%%
\begin{deluxetable}{lcccc}
\tablecolumns{5}
\footnotesize
\tablecaption{The parameters of best-fit models in $\rm DCO^+$ lines}
\tablewidth{0pt}
\tablehead{
\colhead{Source}                  &
\colhead{$X_0$}&
\colhead{$r_D$}&
\colhead{$f_D$}&
\colhead{$V_{\rm microturbulence}$} \\
\colhead{}&
\colhead{}&
\colhead{pc}&
\colhead{}&
\colhead{$\rm km~s^{-1}$}
}
\startdata
 L1512   &  $2.8\times10^{-10}$ &       &    & 0.09\\
 L1544   &  $3.5\times10^{-10}$ & 0.022 &  3 & 0.12\\
 L1689B  &  $3.5\times10^{-10}$ & 0.011 &  4 & 0.20\\
\enddata
\end{deluxetable}

%%%%%%%%%%%%%%%%%%%table10%%%%%%%%%%%%%%%%%%%%%%%%%%%%%%%%%%%%%%%%
\begin{deluxetable}{lcccccccc}
\tablecolumns{5}
\footnotesize
\tablecaption{The parameters of best-fit models in $\rm HCO^+$ lines}
\tablewidth{0pt}
\tablehead{
\colhead{Source}&
\colhead{}&
\colhead{step function}&
\colhead{}&
\colhead{}&
\colhead{}&
\colhead{exponential function\tablenotemark{a}}&
\colhead{}&
\colhead{$V_{\rm microturbulence}$} \\
\colhead{}&
\colhead{$X_0$}&
\colhead{$r_D$}&
\colhead{$f_D$}&
\colhead{}&
\colhead{}&
\colhead{$X_0$}&
\colhead{$r_D$}&
\colhead{} \\
\colhead{}&
\colhead{}&
\colhead{pc}&
\colhead{}&
\colhead{}&
\colhead{}&
\colhead{}&
\colhead{pc}&
\colhead{$\rm km~s^{-1}$}
}
\startdata
 L1512   &  $1.0\times10^{-9}$ &       &    &&&&& 0.1\\
 L1544   &  $3.0\times10^{-9}$ & 0.028 &  10&& &$6.0\times10^{-9}$& 0.04&0.15\\
 L1689B  &  $3.2\times10^{-9}$ & 0.023 &  15&& &$6.5\times10^{-9}$& 0.03&0.20\\
\enddata
\tablenotetext{a}{$X=X_0~exp(-\frac{r_D}{r})$
}
\end{deluxetable}

%%%%%%%%%%%%%%%%%%%table11%%%%%%%%%%%%%%%%%%%%%%%%%%%%%%%%%%%%%%%%
\begin{deluxetable}{lccc}
\tablecolumns{4}
\footnotesize
\tablecaption{The results of this study }
\tablewidth{0pt}
\tablehead{
\colhead{}                  &
\colhead{L1512}&
\colhead{L1544}&
\colhead{L1689B} 
}
\startdata
 Degree of central condensation &  Young & Evolved & Evolved\\
 Chemical evolution (depletion of molecules)   &  Evolved & Evolved & Young\\
 Timescale   &  $\tau_{che}<<\tau_{dyn}$ &$\tau_{che}\approx\tau_{dyn}$ & $\tau_{che}>>\tau_{dyn}$\\
Dynamical state &Stable (?) &AD (?) & Free-fall (?)\\
\enddata
\end{deluxetable}


\begin{references}
\noindent

\reference{}Aikawa, Y., Ohashi, N., Inutsuka, S.-I., Herbst, E. \& Takakuwa, 
S. 2001, ApJ, 552, 639
\reference{}Adams, F.C., Lada, C.J., \& Shu, F.H. 1987, ApJ, 312, 788
\reference{}Andr\'e, P., Ward-Thompson, D., \& Barsony, M. 1993, ApJ, 406, 122
\reference{}Andr\'e, P., Ward-Thompson, D., \& Motte, F., 1996, A\&A, 314, 625

\reference{}Bacmann, A., Lefloch, B., Ceccarelli, C., Castets, A., Steinacker, J.,
\& Loinard, L., 2002, A\&A, 389, L6
\reference{}Bergin, E.A. \& Langer, W.D., 1997, ApJ, 486, 316
\reference{}Bergin, E.A., Alves, J., Huard, T.L., \& Lada, C.J. 2002, 
ApJ, 570, L101
\reference{}Benson, P.J., Caselli, P., \& Myers, P., 1998, 506, 743
\reference{}Blake, G.A., Sandell, G. Van Dishoeck, E.F., Groesbeck, T.D., 
Mundy, L.G., \& Aspin, C., 1995, ApJ, 441, 689
\reference{}Bonnor, W.B. 1956, MNRAS, 116, 351

\reference{}Caselli, P., Walmsley, C. M., Tafalla, M., \& Herbst, E. 1998, ApJ,
499, 234
\reference{}Caselli, P., Walmsley, C.M., Tafalla, M., \& Myers, P.C. 1999, ApJ, 523, L165
\reference{}Caselli, P., Walmsley, D.M., Zucconi, A., Tafalla, M., Dore, L., 
\& Myers, P.C., 2002, ApJ, 565, 331a
\reference{}Caselli, P., Walmsley, D.M., Zucconi, A., Tafalla, M., Dore, L.,
\& Myers, P.C., 2002, ApJ, 565, 344b
\reference{}Caselli, P., 2002, astro-ph/0204127 

\reference{}Ciolek, G.E. \& Mouschovias, T.C., 1994, ApJ, 425, 142
\reference{}Crutcher, R.M. \& Troland, T.H., 2000, ApJ, 537, L139

\reference{}Ebert, R. 1955, Z. Astrophys., 37, 217
\reference{}Evans, N.J.II, Rawlings, J.M.C., Shirley, Y., \& Mundy, L.G.
2001, ApJ, 557, 193
\reference{}Flower, D.R., \& Launay, J.M, 1985, MNRAS, 214, 271
\reference{}Flower, D.R., 1999, MNRAS, 305, 651
\reference{}Foster, P.N., \& Chevalier, R.A., 1993, ApJ, 416, 303
\reference{}Galli, D., Walmsley, M., \& Goncalves, J., 2002, astro-ph/0208416
\reference{}Goldsmith, P.F., 2001, ApJ, 557, 736
\reference{}Gregersen, E.M., Evans, N.J.II., Zhou, S., \& Choi, M., 1997, ApJ, 
484, 256
\reference{}Gregersen, E.M., Evans, N.J.II., 2000, ApJ, 538, 260
\reference{}Jessop, N.E. \& Ward-Thompson, D., 2001, MNRAS, 323, 1025
\reference{}Jorgensen, J.K., Schoier, F.L., \& van Dishoeck, E.F.,  2002, 
A\&A, 389, 908

\reference{}Kramer, C., Alves, J., Lada, C.J., Sievers, A., Ungerechts, H., \&
Walmsley, C.M., 1999, A\&A, 342, 257
\reference{}Kuiper, T.B.H., Langer, W.D., \& Velusamy, T., 1996, ApJ, 468, 761

\reference{}Lada, C.J., Lada, E.A., Clemens, D.P., \& Bally J. 1994,
ApJ, 429, 694
\reference{}Ladd, E.F., Fuller, G.A., \& Deane, J.R., 1998, ApJ, 495, 871
\reference{}Lacy, J.H., Knacke, R., Geballe, T.R., \& Tokunaga ,A.T., 1994,
ApJ, 428, L69
\reference{}Larson, R.B., 1969, MNRAS, 145, 271
\reference{}Lee, C.W., Myers, P.C., \& Tafalla, M., 1999, ApJ, 526, 788
\reference{}Li, Z.-Y., Shematovich, V.I., Wiebe, D.S., \& Shustov, B.M., 2002,
ApJ, ApJ, 569, 792
\reference{}Loren, R.B., 1989, ApJ, 338, 902

\reference{}McMullin, J.P., \& Mundy, L.G., 1994, ApJ, 424, 222
\reference{}McLaughlin, D.E., \& Pudritz, R.E., 1997, ApJ, 476, 750
\reference{}Myers, P.C. \& Ladd, E.F., 1993, ApJ, 413, L47
\reference{}Myers, P.C., Mardones, D., Tafalla, M., Williams, J.P., \& Wilner,
D.J. 1996, ApJ, 465, L133

\reference{}Ohashi, N., Lee, S.W., Wilner, D.J., \& Hayashi, M., 1999, ApJ, 518, L41
\reference{}Ossenkopf V. \& Henning Th., 1994, A\&A, 291, 943

\reference{}Penston, M.V., 1969, MNRAS, 144, 425
\reference{}Ruffle, D.P., Hartquist, T.W., Taylor, S.D., \& Williams, D.A., 
1997, MNRAS, 291, 235
\reference{}Ruffle, D.P, Hartquist, T.W., Caselli, P., Rawlings, J.M.C. \&
Williams, D.A., 1999, Astrophyscis and Space Science, 262, 177
\reference{}Rawlings, J.M.C., Hartquist, T.W., Menten, K.M., \& Williams, D.A., 
1992, MNRAS, 255, 471
\reference{}Rawlings, J.M.C., 2000, IAU symposium, vol. 197, 15
\reference{}Rawlings, J.M.C. \& Yates, J.A., 2001, MNRAS, 326, 1423 

\reference{}Shirley, Y.L., Evans, N.J.II, Rawlings, J.M.C. \& Gregersen, E.M. 
2000, ApJS, 131, 249
\reference{}Shu, F.H. 1977, ApJ, 214, 488
\reference{}Shu, F.H., Adams, F. C., \& Lizano, S. 1987, ARAA, 25, 23
\reference{}Suzuki, H., Yamamoto, S., Ohishi, M., Kaifu, N., Ishikawa, S., 
Hirahara, Y., \& Takano, S., 1992, ApJ, 392, 551

\reference{}Tafalla, M., Mardones, D., Myers, P.C., Caselli, P., Bachiller, R., 
\& Benson, P.J., 1998, ApJ, 504, 900
\reference{}Tafalla, M., Myers, P.C., Caselli, P., Walmsley, C.M., \& Comito, 
C., 2002, ApJ, 568, 815
\reference{}Troland, T.H., Crutcher, R.M., Goodman, A.A., Heiles, C., Kazes, I.,
\& Myers, P.C., 1996, ApJ, 471, 302 

\reference{}van Dishoek, E.F., Blake, G.A., Draine, B.T., \& Lunine, J.I, 1993,
in Protostars and Planets III, Levy E.H. \& Lunine J.I. (eds.). Tucson: 
University of Arizona Press, p. 163  
\reference{}Ward-Thompson, D., Scott, P.E., \& Andr\'e, P., 1994, MNRAS, 268, 276
\reference{}Ward-Thompson, D., Motte, F., \& Andr\'e, P., 1999, MNRAS, 305, 143
\reference{}Ward-Thompson, D., Kirk, J.M., Crutcher, R.M., Greaves, J.S., 
Holland, W.S., \& Andr\'e, P., 2000, ApJ, 537, L135
\reference{}Ward-Thompson, D., Andr\'e, P., \& Kirk, J.M., 2002, MNRAS, 329, 257
\reference{}Watson, W.D. 1977, CNO Isotopes in Astrophysics, ed. J. Audouze,
Reidel Publ. (Dordrecht)
\reference{}Whitworth, A.P. \& Ward-Thompson, D., 2001, ApJ, 547, 317
\reference{}Willacy, K., Langer, W.D., \& Velusamy, T., 1998, ApJ, 507, L171 
\reference{}Williams, J.P., Myers, P.C., Wilner, D.J., \& di Francesco, J., 
1999, ApJ, 513, L61
\reference{}Wilson, T.L. \& Rood, R.T., 1944, ARAA, 32, 191 

\reference{}Zhou, S. Evans, N.J.II, Kompe, C., \& Walmsley, C. M. 1993, ApJ,
404, 232
\reference{}Zucconi, A., Walmsley, C.M., \& Galli, D., 2001, A\&A, 376, 650

\end{references}
\end{document}